\documentstyle[psfig]{mn}

\newif\ifAMStwofonts
\AMStwofontstrue

\def\Ha{\hbox{$\hbox{H}_\alpha$}}

\def\lsim{\mathrel{\lower2.5pt\vbox{\lineskip=0pt\baselineskip=0pt
           \hbox{$<$}\hbox{$\sim$}}}}
\def\gsim{\mathrel{\lower2.5pt\vbox{\lineskip=0pt\baselineskip=0pt
           \hbox{$>$}\hbox{$\sim$}}}}

\ifoldfss
  \ifCUPmtlplainloaded \else
    \NewTextAlphabet{textbfit} {cmbxti10} {}
    \NewTextAlphabet{textbfss} {cmssbx10} {}

    \NewMathAlphabet{mathbfit} {cmbxti10} {} 
    \NewMathAlphabet{mathbfss} {cmssbx10} {} 
  \fi
  \ifAMStwofonts
    \ifCUPmtlplainloaded \else
      \NewSymbolFont{upmath} {eurm10}
      \NewSymbolFont{AMSa} {msam10}
      \NewMathSymbol{\upi}     {0}{upmath}{19}
      \NewMathSymbol{\umu}     {0}{upmath}{16}
      \NewMathSymbol{\upartial}{0}{upmath}{40}
      \NewMathSymbol{\leqslant}{3}{AMSa}{36}
      \NewMathSymbol{\geqslant}{3}{AMSa}{3E}

      \let\leq=\leqslant 
      \let\geq=\geqslant 
    \fi
  \fi
\fi 

\ifnfssone
  \newmathalphabet{\mathit}
  \addtoversion{normal}{\mathit}{cmr}{m}{it}
  \addtoversion{bold}{\mathit}{cmr}{bx}{it}
  \newmathalphabet{\mathbfit} 
  \addtoversion{normal}{\mathbfit}{cmr}{bx}{it}
  \addtoversion{bold}{\mathbfit}{cmr}{bx}{it}
  \newmathalphabet{\mathbfss} 
  \addtoversion{normal}{\mathbfss}{cmss}{bx}{n}
  \addtoversion{bold}{\mathbfss}{cmss}{bx}{n}
  \ifAMStwofonts
    \ifCUPmtlplainloaded \else
      \UseAMStwoboldmath
      \makeatletter
      \new@mathgroup\upmath@group
      \define@mathgroup\mv@normal\upmath@group{eur}{m}{n}
      \define@mathgroup\mv@bold\upmath@group{eur}{b}{n}
      \edef\UPM{\hexnumber\upmath@group}
      \new@mathgroup\amsa@group
      \define@mathgroup\mv@normal\amsa@group{msa}{m}{n}
      \define@mathgroup\mv@bold\amsa@group{msa}{m}{n}
      \edef\AMSa{\hexnumber\amsa@group}
      \makeatother
      \mathchardef\upi="0\UPM19
      \mathchardef\umu="0\UPM16
      \mathchardef\upartial="0\UPM40
      \mathchardef\leqslant="3\AMSa36
      \mathchardef\geqslant="3\AMSa3E

      \let\leq=\leqslant 
      \let\geq=\geqslant 
    \fi
  \fi
\fi 

\ifnfsstwo
  \DeclareMathAlphabet{\mathbfit}{OT1}{cmr}{bx}{it}
  \SetMathAlphabet\mathbfit{bold}{OT1}{cmr}{bx}{it}
  \DeclareMathAlphabet{\mathbfss}{OT1}{cmss}{bx}{n}
  \SetMathAlphabet\mathbfss{bold}{OT1}{cmss}{bx}{n}
  \ifAMStwofonts
    \ifCUPmtlplainloaded \else
      \DeclareSymbolFont{UPM}{U}{eur}{m}{n}
      \SetSymbolFont{UPM}{bold}{U}{eur}{b}{n}
      \DeclareSymbolFont{AMSa}{U}{msa}{m}{n}
      \DeclareMathSymbol{\upi}{0}{UPM}{"19}
      \DeclareMathSymbol{\umu}{0}{UPM}{"16}
      \DeclareMathSymbol{\upartial}{0}{UPM}{"40}
      \DeclareMathSymbol{\leqslant}{3}{AMSa}{"36}
      \DeclareMathSymbol{\geqslant}{3}{AMSa}{"3E}

      \let\leq=\leqslant 
      \let\geq=\geqslant 
    \fi
  \fi
\fi 

\ifCUPmtlplainloaded \else
  \ifAMStwofonts \else 
    \def\upi{\pi}
    \def\umu{\mu}
    \def\upartial{\partial}
  \fi
\fi

\title{ Structural parameters of nearby emission-line galaxies }

\author [Miguel S\'anchez-Portal et al.]
       {Miguel S\'anchez-Portal$^1$ \thanks{e-mail: miguel.sanchez@upsam.net
       },  
\'Angeles I. D\'{\i}az$^2$,  
\newauthor Elena Terlevich$^3$\thanks{Visiting Fellow, IoA,
       Cambridge},  Roberto Terlevich$^{3,4}$ \\
$^1$ Universidad Pontificia de Salamanca en Madrid,
Paseo de Juan XXIII 3, 28040 Madrid, Spain\\
$^2$ Dpto. de F\'{\i}sica Te\'orica, C-XI, 
Universidad Aut\'onoma de Madrid, Cantoblanco, 28049 Madrid,
Spain\\                                                             
$^3$ Instituto Nacional de Astrof\'{\i}sica, \'Optica y Electr\'onica, 
Tonantzintla, Puebla, M\'exico\\
$^4$ Institute of Astronomy, Madingley Road, Cambridge, CB3 0HA, U.K. }
\date{}

\pagerange{\pageref{firstpage}--\pageref{lastpage}}
\pubyear{2003}

\begin{document}

\maketitle

\label{firstpage}

\begin{abstract}
We present the results of an investigation on the main structural
properties derived from {\em VRI} and \Ha\ surface photometry of galaxies hosting
nuclear emission-line regions (including Seyfert~1, Seyfert~2, LINER and
starburst galaxies) as compared with normal galaxies. Our original sample
comprises 22 active galaxies, 4 starbursts and 1 normal galaxy and has been
extended with several samples obtained from the literature. Bulge and disc
parameters, along with $B/D$ relation, have been derived applying an iterative
procedure. The resulting parameters have been combined with additional data in
order to reach a statistically significant sample. We find some differences
in the bulge distribution across the different nuclear types that could
imply familes of bulges with different physical properties. Bulge and disc
characteristic colours have been defined and derived for our sample and
compared with a control sample of early type objects. The results suggest that
bulge and disc stellar populations are comparable in normal and active galaxies.

\end{abstract}

\begin{keywords}
galaxies: active -- galaxies: surface photometry -- galaxies: morphology
\end{keywords}

\section{Introduction}

Global AGN host galaxy properties, including mass, luminosity concentration,
morphological type, bulge-to-disc relationship, metal abundance and
total luminosity, may influence nuclear activity and perhaps in turn
be affected by it. 
It has been argued that there can exist a connection between
nuclear activity and intense star formation episodes. While there are
some observational indications that star formation is enhanced in
circumnuclear regions of Seyfert galaxies (see for instance Hunt \&
Giovanardi 1992), other observations point in
the opposite sense (e.g. Carone 1992).
An outstanding, still controversial question is related to the fuel supply
involving the transport of large quantities of mass from reservoirs
located in the galactic disc (see for instance Shlosman et al. 1989); 
in this process the matter 
has to lose almost all of its angular momentum. This can be
accomplished by gravitational torques produced by non-axisymmetric
perturbations in the gravitational potential. These are associated either
with interactions with companion galaxies 
or with structural features like bars, rings, oval distortions etc.  

Current research on structural properties of galaxies hosting nuclear 
activity is now largely concentrated in  the near infrared (NIR) bands. 
There are several advantages in the use of the NIR for studying the
host galaxy properties of AGNs: on the one hand, a normal   stellar
population is dominated by old red giants and therefore peaks around
1$\mu$m. On the other, dust extinction is substantially reduced at
these wavelengths. Therefore, the NIR is ideal to look for bars and 
morphological distortions that
can be hidden by dust in optical wavelengths (e.g. Mulchaey \& Regan
1997). Moreover, the contrast between stellar and non-thermal
contributions is maximised, since the latter has a local minimum
around 1$\mu$m.

Hunt et al. \shortcite{Hunt1997} performed NIR $J$, $H$, $K$ broadband
images and colours in a sample of 26 galaxies with $ z \geq 0.015$ and
found that the average colours, both of the outer and the inner discs,
are independent of the activity class and equal to those of normal
spiral galaxies. Outer disc colour gradients are also consistent with
those of normal spirals. On the other hand, on a scale of a few
kiloparsecs from the centre, starburst galaxies show ongoing star
formation; in Seyfert~2 galaxies, they found evidence of
``fossil'' star formation activity a few hundred Myr old; finally, no
evidence of star formation, either ongoing or fossil, exceeding that
of a normal spiral, is found in Seyfert~1 galaxies. They infer an
evolutionary starburst-Seyfert connection, although not so global and not so
tight in time as previous studies implied. NIR structural 
parameter analysis from Hunt et al. \shortcite{Hunt1999} shows that bulges of
Seyfert and starburst galaxies follow very closely the trend defined by normal
spiral bulges and elliptical galaxies. On the other hand, for a given disc
effective radius, the mean surface brightness of the Seyfert discs is higher
(by $\sim$ 0.9 K mag arcsec$^{-2}$) than those of early-type spirals. 

M\' arquez et al \shortcite{Marquez2000} performed NIR $J$ and $K'$ broadband
imaging and colours on nearby, isolated spiral galaxies either 
 with an active nucleus (18 objects, either Seyfert~1 or Seyfert~2
 according to V\' eron-Cetty \& V\' eron 1993) or without it (11
 galaxies). They state that global properties are similar in both active and
 non active galaxies, since both samples define the same bulge Kormendy relation
 between $\mu_e$ and $r_e$\footnote{Nevertheless, there is a difference in the
   slope of the Kormendy relation of active and non active galaxies at $\sim 2
   \sigma$ level that is not considered as relevant.}
 and disc components also share the same properties
 (contradicting the result from Hunt 1999); they also found that bulge and
disc scale lengths are correlated, with $r_{bulge} \approx 0.2 r_{disc}$. On
the other hand, NIR central colours of active galaxies are found to be 
redder than
 those of non active spirals, most probably due to the AGN light re-emitted by
 the hot dust and/or to the presence of active star formation in circumnuclear
 regions.

In spite of the advantages of the NIR region for such studies, optical wavelengths can
still provide useful data; the use of the optical region has its own advantages: on the one hand, 
observers can benefit from the large size of current optical detectors, providing a 
convenient sampling
of the galaxy disc; on the other, optical studies can rely on the large quantity of published data.
Moreover, optical colour maps can be used as excellent tracers of the line emission regions 
present in AGNs and starburst galaxies (see, for instance Terlevich et al. 1991)

Early work from Yee \shortcite{ApJ272:Yee} using a SIT-vidicon camera has been
followed by a large number of studies using CCD detectors. Mediavilla et
al. (1989) and  Pastoriza et al. \shortcite{Pastoriza} performed {\em VI} 
broadband surface
photometry on a sample of 13 Seyfert galaxies; they consider two classes:
``bulge galaxies'' and ``disc galaxies'' and found that the  
 \mbox{\em V-I} colour has a constant, equivalent value in both galaxy groups
 in the 3~Kpc aperture, indicating similar stellar populations, but outside 
the 5~Kpc aperture colours become bluer for disc galaxies and redder for bulge
galaxies.  Mediavilla et al. \shortcite{ApSS157:Mediavilla} use the same
sample to obtain bulge and disc parameters from the surface brightness radial
profiles, finding that structural parameters are comparable to those of normal
spirals and ellipticals.

MacKenty \shortcite{ApJS72:MacKenty} performed Kron-Cousins {\em BVR} CCD surface
photometry of 51 Seyfert galaxies from the NGC and Markarian catalogs finding
that in almost all cases, a mechanism to transport material to the nucleus is
present (either bars or interactions); about one-fifth of the sample is
composed by amorphous galaxies; some of them can be a remnant of past
interactions. Colours and disc morphological parametrs are in general similar
to those found in normal spiral galaxies. 

Xanthopoulos \shortcite{MNRAS280:Xanthopoulos} performed {\em VRI} CCD surface
photometry of 27 Seyfert galaxies belonging to type 1, 2 and intermediate
categories, finding that:
{\em (a)} generally, Seyfert galaxies are perturbed objects (either by bars or
by interactions); {\em (b)} nearly half of the sample galaxies are barred;
moreover, almost all of them present evidence of the existence of a mechanism
to transfer material to the nuclear region;  {\em (c)} all non-barred galaxies
are either S0 or Sa; {\em (d)} Seyfert~1 and Seyfert~2 galaxies follow a very
similar Hubble type distribution; Seyfert~2 galaxies tend to have redder
nuclei; {\em (e)} Seyfert~1 and Seyfert~2 galaxies present the same disc colours.
{\em (f)} colour gradients remain constant along  both the bulge and the disc
of the galaxy for both Seyfert types; {\em (g)} both types of Seyfert galaxy
show the same range and dispersion of bulge and disc morphological parameters,
and also similar bulge-to-disc luminosity ratios; the wide range of scale
lengths observed in Seyfert galaxies may support their classification as
early-type spirals. Moreover, the bulge and disc parameters of both Seyfert
types are the same as those of normal early-type spirals.

Virani et al. \shortcite{Virani} performed {\em R} surface photometry on
a sample of 32 Seyfert and 45 non active galaxies\footnote{From the
control sample, at least two galaxies have been found to be active and
at least five objects are nuclear starbursts according to 
Ho et al. \shortcite{Ho1997}.} from the CfA
survey and derived bulge and disc morphological parameters from the
azimuthally averaged surface brightness profiles. 
No statistically significant differences between
the Seyfert and control samples are found in either morphological
parameters or in light asymmetries (bars, rings, isophotal twisting,
etc.), or even in the companion galaxies properties (magnitude,
separation from the host, position angle relative to the host,
magnitude difference  between the companion and the host, strength of
the tidal parameter). They conclude that the nearby environment of
Seyfert galaxies is not significantly different from nonactive
galaxies with the same morphological distribution. 

Boris et al. \shortcite{Boris} performed {\em BVI} surface photometry
on a sample of 10 Seyfert~1 galaxies along with narrow-band \Ha\
images of a subsample of 6 objects. They found that morphologies are
confined to early-type galaxies; half of them can be considered as
compact; bars are found in only two cases; average \mbox{\em B-V}
colours are found bluer than those expected for this morphological
type (perhaps due to the contribution of the active nucleus and/or the
disc to the total luminosity). Evidence of tidal interactions is
found in six objects of the sample. They obtain morphological
parameters for the nucleus (gaussian), bulge and disc components
finding that in most cases (six out of eight), the disc is best described
by a truncated  exponential profile. The cutoff is generally associated with
the existence of reddened regions (probably dust in the inner few Kpc
of the galaxy). Only one galaxy shows disc emission in \Ha .

This paper presents a new optical study on the structural parameters 
of galaxies hosting nuclear
emission-line regions. In order to overcome the usual weakness of this
kind of studies (relatively small and frequently biased samples, 
uncertainties in the nuclear
classification, specially in the control samples), 
whenever possible, we have tried to supplement our
original sample with data from the literature, in particular morphological parameters,
and to use an homogeneous,
reliable nuclear classification from optical spectroscopy. When data from the literature 
is used, it is clearly indicated throughout.  The paper
is structured as follows: first, a description of the original sample
and the adopted nuclear classification is outlined; second, a description of the
bulge and disc decomposition procedure is given, followed by the
results obtained; the discussion section then proceeds, including a thorough analysis
of the structural parameters of galaxies hosting nuclear emission-line
regions. 
Our analysis is closed with a discussion on the radial
colour distribution of our sample objects. Finally, our conclusions
are outlined in section \ref{conclusions}.

\section{The sample}

Our sample (described in S\'anchez-Portal et al. 2000, hereafter Paper I)
comprises 27 nearby galaxies in a wide range of Hubble types 
(from $T = -5$ to $T = 10$) and different level of nuclear activity. 

The availability of an accurate and homogeneous  nuclear
classification is of critical importance  when trying to determine the
variation of the behaviour of 
a given physical property across a sample of galaxies with different degree
of activity.  To this end, in this paper we have modified the original nuclear
classification from Paper I and  adopted the nuclear types
from Ho et al. \shortcite{Ho1997} (hereafter HFS), since it provides a 
reliable 
homogeneous framework. In addition to \mbox{H{\small II}} nuclei,
Seyferts and LINERs, HFS recognises a group of ``transition objects'',
whose \mbox{[O{\small I}]} strengths are intermediate between those of
\mbox{H{\small II}} nuclei and LINERs; transition objects can be most 
naturally explained as ``normal'' LINERs whose integrated spectra are 
diluted or contaminated by neighbouring \mbox{H{\small II}}
regions. The HFS survey demonstrates that broad \Ha\ exists not only in
Seyfert~1 nuclei but also in LINERs and perhaps even in some
transition objects. Accordingly, they propose to extend the ``type 1''
and ``type 2'' designations, along with the intermediate types (1.2,
1.5, 1.8 and 1.9) to LINERs and transition objects.
Table \ref{new_types} shows the
old and current nuclear types. According to the adopted 
HFS classification, several nuclear types are 
different from the original ones presented in Paper I.
Some objects experience a major
modification: NGC~3998 and NGC~5077, formerly considered Seyfert
galaxies, become LINERs, while M~106 and M~51, previously classified
as LINERs are now included in the Seyfert category. NGC~6340 and
NGC~6384, initially considered normal galaxies, fall now in the LINER
class. Nevertheless, the main results outlined in Paper I are not
affected.

\begin{table}
\centering
\begin{minipage}{60mm}
\caption{Original and revised nuclear types. 
\newline
S = Seyfert, L = LINER, T
= transition nucleus, SB = \mbox{H{\small II}} nucleus.}
\label{new_types}
\begin{tabular}{lcc}
\hline
Galaxy & Old type & New type\\
\hline
NGC 3227 & S1.5 & S1.5 \\ 
NGC 3516 & S1.5 & S1.2 \\ 
NGC 3998 & S1.5 & L1.9 \\ 
NGC 7469 & S1 & S1\footnote{Not present in the HFS catalog}\\
NGC 4151 & S1.5 & S1.5 \\ 
NGC 5077 & S1 & L1.9 \\ 
NGC 6814 & S1 & S1$^{a}$ \\ 
NGC 513  & S2 & S2$^{a}$ \\
NGC 1068 & S2 & S1.8 \\
Mrk 620 (NGC 2273) & S2 & S2 \\ 
Mrk 622  & S2 & S2$^{a}$ \\
NGC 3982 & S2 & S1.9 \\ 
NGC 5347 & S2 &  S2$^{a}$ \\
NGC 6217 & S2 &  SB \\ 
NGC 7479 & S2 &  S1.9 \\ 
NGC 1052 & L &  L1.9 \\
NGC 2841 & L &  L2 \\
M 106 (NGC 4258) & L  & S1.9 \\  
M 51 (NGC 5194) & L & S2 \\
NGC 7177  & L & T2 \\
NGC 7217 &  L & L2 \\ 
NGC 2146 & SB & SB \\
NGC 3310 & SB & SB \\ 
NGC 3353 & SB & SB$^{a}$ \\ 
NGC 1023 & Normal & Normal  \\ 
NGC 6340 & Normal & L2 \\ 
NGC 6384 & Normal & T2 \\ 
\hline
\end{tabular}
\end{minipage}
\end{table}

\section{Bulge-to-disc decomposition procedure}

A spiral or lenticular galaxy can be decomposed to a first
approximation into a bulge and a disc. Despite several differences
(e.g. greater rotational speed), bulges are generally considered as
equivalent to elliptical galaxies due to their morphological
similarity, surface brightness distribution and stellar content
\cite{lectures:Kormendy}. Therefore, the same analytical functions are used for
ellipticals and bulges of spiral and S0 galaxies. Perhaps the most
widely used fitting function is that from de~Vaucouleurs
\shortcite{deVauc:1948}:

\begin{equation}
I(r) = I_{e} 10^{-3.33[(r/r_{e})^{1/4} -1]}
\label{eq:bulge}
\end{equation}

\noindent
where  $I_e$ is the intensity at $r_e$, the effective radius. 
The factor 3.33 is chosen so that half of the total light emitted by
 the model comes from inside $r_e$.

On the other hand, the disc component can be represented by an
exponential law \cite{ApJ160:Freeman}:

\begin{equation}
I(r) = I_{0}e ^{(-r/r_0)}
\label{eq:disk}
\end{equation}

\noindent
where $I_0$ is the disk central surface brightness and
$r_0$ the disc scale length. Though the reason for an exponential
surface brightness distribution is uncertain, it can be derived from a
possible solution of the gravitational collapse of a uniform density
rotating sphere (see for example, Freeman 1970, Kormendy 1983). 
Freeman  \shortcite{ApJ160:Freeman} distinguishes two types of
galactic discs:
{\em Type I} discs reach $r=0$ while 
{\em Type II} do not, showing an inner cutoff. This can be
due to an actual lack of mass with low angular momentum, perhaps in the
protogalactic cloud, thus leaving a galactic disc with an inner
``hole''. Alternatively, type II discs  can be the result of
the combination of a pure exponential disc plus an additional component
like a lens \cite{lectures:Freeman}. Kormendy
\shortcite{ApJ217:Kormendy} proposed a formula to fit this kind of
discs with internal cutoff, modifying the Freeman's expression (eq.
\ref{eq:disk})

\begin{equation}
I(r) = I_{0}e ^{-[r/r_{0} + (r_{cut} / r)^{\beta}]}
\label{eq:diskcut}
\end{equation}

\noindent
where usually $ \beta = 3 $.

Andredakis et~al. \shortcite{Andredakis95} have fitted the
bulge profile with a generalised exponential profile of S\' ersic
\shortcite{Sersic68}:

\begin{equation}
I(r) = I_{0}e ^{-(r/r_{0})^{1/n}}
\label{eq:Sersic}
\end{equation}

\noindent
finding that the exponent {\em n} varies with the Hubble type; while
for early-type spirals, $n=4$ (standard de~Vaucouleurs profile), late-type
galaxies show exponential bulges (i.e. $n=1$). Carollo
et~al. \shortcite{Carollo98} performed the analysis of HST WFPC2 F606W images
of a sample of 75 spiral galaxies, ranging from Sa to Sbc. Only a fraction
($\simeq$ 40\%) of the objects contains a smooth, classical $R^{1/4}$ bulge. The central
structure of some sample objects presents an irregular morphology: it
is often dwarf-irregular like; in a few cases is elongated, similar to
a late-type bar. In several objects, the spiral structure reaches down
to the innermost accessible scales. Ocasionally, a small
``bulge-like'' feature coexists with the nuclear spiral
structure. They found that, in several cases, these inner,
morphologically-distinct structures are well fitted by an exponential
profile. These exponential bulges embedded in the spiral
structure appear to be fainter than $R^{1/4}$ bulges for a given total
galactic luminosity and Hubble type.

Two basic methods have been customarily used for obtaining the bulge and disc
components from the azimuthally-averaged surface brightness profiles:
non-linear least squares fitting and iterative 
fitting. In the former method, a non-linear least squares
algorithm is used in order to fit simultaneously both components, while
in the latter, an iterative method is applied in order to fit both
components independently: a linear least squares fit is applied first
to the region dominated by one of the components; this first model is
extrapolated to the whole range, and subtracted from the observed
profile, giving a first estimate of the other component, and then the
appropriate fitting law is applied, subtracting the obtained model
from the observed profile. This residual is then fitted again and the
procedure is repeated iteratively until convergence is achieved. The
regions where each component dominates over the other should be
estimated by visual inspection. Schombert \& Bothum
\shortcite{AJ:Schombert} propose a combination of both procedures:
first, an iterative procedure is used until convergence is
achieved. The resulting bulge and disc parameters are then used as
initial estimates for a non-linear least squares fitting procedure.
de Jong \shortcite{deJongII} has applied a two-dimensional
method using two or three 2D components to model the bulge, disc and (whenever
appropriate) bar and applying a non-linear algorithm capable of accepting
different weights for each data point. An exponential bulge is
proposed instead of the classical $R^{1/4}$ model.

In spite of the disadvantage of the subjective step of visual
determination of  the ranges of dominance of disc or bulge, we have chosen
the iterative fitting procedure because it allows to control the
evolution of the procedure and to exclude regions from the fitting if
desired (e.g. regions dominated by a ring or lens, where the profile
is not properly described by the sum of a bulge plus a disc). On the
other hand, it is possible to detect a disc cutoff
(i.e., type II 
discs). We have used as independent variable of the surface brightness
profiles the equivalent radius,
defined as $r_{eq} = \sqrt{A/\pi}$, where {\em A} is the area enclosed
by a given isophote. The most accurate fittings are obtained when there
exists a clear range of dominance of one component over the other,
i.e. when bulge and disc scale lengths are rather different. In those
cases, the fitting has been always started by determining the disc
dominance region. On the other hand, for those galaxies where a disc
component never dominates, the fitting has been started by a bulge
component. In those latter cases the fitting confidence depends on the
accuracy of the $R^{1/4}$ law defining the bulge
\cite{ApJS59:Kent}. We have also computed the  integrated
bulge-to-disc luminosity ratio:

\begin{equation}
B/D  \equiv \frac{ \int_0^{\infty} I_{bulge}(r)2 \pi rdr}{
  \int_0^{\infty} I_{disc}(r)2 \pi rdr} = 3.61 
\left(\frac{r_e}{r_0}\right)^2 \frac{I_e}{I_0}
\label{bd_1}
\end{equation} 

\noindent
where $I_{bulge}(r)$ and $I_{disc}(r)$ are the surface brightness
distributions for bulge and disc given by equations \ref{eq:bulge} and
\ref{eq:disk}. If the galaxy has a type II disc, equation \ref{bd_1}
is no longer valid. Instead of using a fitting formula like
equation \ref{eq:diskcut} for this kind of discs, we have simply assumed that
the galaxy has a sharp 
cutoff for certain radius $r = r_{cut}$. Therefore, we change the
lower limit of the disc luminosity integral in equation \ref{bd_1},
obtaining the expression:

\begin{equation}
B/D  = 3.61 \left(\frac{r_e}{r_0}\right)^2 \frac{I_e}{I_0}
\frac{e^{r_{cut}/r_0}}{\left(\frac{r_{cut}}{r_0} + 1 \right)}
\label{bd_2}
\end{equation} 

Our computation method assumes that, beyond certain distance from the
galaxy centre, the effect of seeing on the surface brightness profile
is negligible, and therefore the derived bulge and disc parameters are
seeing-free. This assumption is supported by the calculations
performed by Capaccioli \& de Vaucouleurs \shortcite{Capaccioli83}; 
they plotted the convolution function
of the light distribution of an idealized E0 galaxy obeying the
$R^{1/4}$ with a single gaussian seeing PSF, for different values of $r_e$
and $\sigma_{PSF}$, and shown that there is a radius
$r_{ini}(r_e, \sigma_{PSF})$ where the difference between the $R^{1/4}$
model and its convolved image becomes negligible. We have estimated
$r_{ini}$ using these curves and an  initial guess of $r_e$
and $\sigma_{PSF}$. Furthermore, we have tried to ensure that our computation
is not contaminated by nuclear emission by checking the radial
extension of the
nuclear \Ha\ emitting regions from our narrow-band surface photometry (see
Paper~I) and using as initial radius for profile fitting the maximum
between $r_{ini}$ and the radial emission extension.

\section{Results}
 
Table \ref{param_morfo} 
presents the resulting morphological parameters
for the sample galaxies. Magnitudes have been corrected for
inclination using the values from 
Paper~I\footnote {H$_0$= 55 km s$^{-1}$ Mpc$^{-1}$ is used throughout the paper.}. 
The computation of profile data
point errors is detailed in Paper~I. Parameter errors presented here are those arising from
the linear fitting procedure. Figure \ref{fig:parametros1} 
shows the results of the bulge-disc decomposition in the $I$
band. The whole set of surface brightness profiles is presented in Paper~I. 

\begin{table*}
\label{param_morfo}
\begin{center}
\caption{Morphological parameters of the sample galaxies}
\end{center}
\begin{tabular}{lcccccc}
\hline 
Galaxy & Filter & $r_{e}$ & $\mu_{e}$ & $r_{0}$ &
  $\mu_{0}$  & {\it B/D}\\
 & & (\arcsec )  & (mag/\arcsec$^{2}$) & (\arcsec ) & (mag/\arcsec$^{2}$) & \\
 & & (Kpc) &  & (Kpc) &  & \\
\hline 
NGC 3227 &  V & 1.614 $\pm$ 0.136  & 18.267 $\pm$ 0.223 & 23.317 $\pm$
  0.630 & 20.56 $\pm$ 0.047 & 0.143 \\ 
&  & 0.165 $\pm$ 0.014 & & 2.378 $\pm$ 0.064 &  &  \\ 
 & R & 1.745 $\pm$ 0.144  & 17.783 $\pm$ 0.217 & 23.313 $\pm$
 0.671 & 20.107 $\pm$ 0.047 & 0.172  \\ 
    &    &  0.178 $\pm$ 0.015 & &  2.378 $\pm$ 0.068 & &  \\  
 & I  & 2.601 $\pm$ 0.070  &  18.144 $\pm$ 0.069 & 23.404 $\pm$
0.511 & 19.546 $\pm$ 0.045 & 0.162 \\
   &   & 0.265 $\pm$ 0.007 & &  2.387 $\pm$ 0.052 & &  \\ 
\\
NGC 3516 & V & 8.948 $\pm$ 0.295  & 20.041 $\pm$ 0.066 & 21.447 $\pm$
  0.439 & 21.940 $\pm$ 0.036 & 5.354 \\ 
  & & 2.070 $\pm$ 0.068 & & 4.962 $\pm$ 0.102 &  &  \\ 
 & R & 7.472 $\pm$ 0.217  & 19.269 $\pm$ 0.062 & 24.836 $\pm$
 1.385 & 21.875 $\pm$ 0.074 & 4.911  \\ 
   &    &  1.729 $\pm$ 0.050 & &  5.747 $\pm$ 0.081 & &  \\   
  & I & 8.491 $\pm$ 0.179  & 18.892 $\pm$ 0.043 & 18.314 $\pm$
  2.217 & 20.906 $\pm$ 0.219 & 8.212  \\
    &  & 1.965 $\pm$ 0.041 & &  4.238 $\pm$ 0.513 & &  \\
\\
NGC 3982 & V & 27.202 $\pm$ 1.417 & 22.512 $\pm$ 0.065 & 8.303 $\pm$
  0.266 & 18.631 $\pm$ 0.106 & 1.884 \\ 
  & & 2.662 $\pm$ 0.139 &  & 0.812 $\pm$ 0.026  &  &  \\
  & R & 21.741 $\pm$ 0.962  & 21.748 $\pm$ 0.058 & 8.288 $\pm$
 0.198 & 18.221 $\pm$ 0.077 & 1.677 \\ 
    &    &  2.127 $\pm$ 0.094 & & 0.810 $\pm$ 0.019 & &  \\   
  & I & 19.239 $\pm$ 0.189  & 20.963 $\pm$ 0.013 & 8.481 $\pm$
  0.241 & 17.874 $\pm$ 0.087 & 1.841  \\
   &   & 1.882 $\pm$ 0.018 & &  0.830 $\pm$ 0.023 & &  \\
\\
NGC 7469 & V & 2.062 $\pm$ 0.057 & 18.097 $\pm$ 0.065 & 30.772 $\pm$
  3.025 & 21.508 $\pm$ 0.102 & 0.375 \\ 
  & & 0.882 $\pm$ 0.024 &  & 13.161 $\pm$ 1.294  &  &  \\
 & R & 1.635 $\pm$ 0.029  & 16.904 $\pm$ 0.044 & 33.873 $\pm$
 7.090 & 21.652 $\pm$ 0.178 & 0.667 \\ 
   &     &  0.699 $\pm$ 0.012 & & 14.487 $\pm$ 3.032 & &  \\ 
 & I & 1.609 $\pm$ 0.057  & 16.332 $\pm$ 0.088 & 15.399 $\pm$
  2.102 & 20.222 $\pm$ 0.239 & 1.419  \\
  &    & 0.688 $\pm$ 0.024 & &  6.586 $\pm$ 0.899 & &  \\ 
\\
NGC 4151 & V & 10.320 $\pm$ 0.987 & 20.001 $\pm$ 0.155 & 39.771 $\pm$
  2.080 & 20.985 $\pm$ 0.035 & 0.602 \\ 
  & & 0.907 $\pm$ 0.087 &  & 3.495 $\pm$ 0.183  &  &  \\
 & R & 6.100 $\pm$ 0.346  & 18.703 $\pm$ 0.102 & 27.596 $\pm$
 1.176 & 20.189 $\pm$ 0.043 & 0.693  \\ 
   &     &  0.536 $\pm$ 0.030 & &  2.425 $\pm$ 0.103 & &  \\  
 & I & 9.852 $\pm$ 0.222  & 18.921 $\pm$ 0.035 & 40.326 $\pm$
  2.155 & 19.947 $\pm$ 0.037 & 0.554  \\
   &   & 0.866 $\pm$ 0.019 & &  3.544 $\pm$ 0.189 & &  \\
\\
M 106 & V  & 12.732 $\pm$ 0.627 & 20.235 $\pm$ 0.079 & 62.609 $\pm$
  9.662 & 20.163 $\pm$ 0.114 & 0.140 \\ 
(NGC 4258) &  & 0.503 $\pm$ 0.025 &  & 2.472 $\pm$ 0.381  &  &  \\
 & R & 13.777 $\pm$ 0.609 & 19.966  $\pm$ 0.070 & 63.880 $\pm$
 7.346 & 19.783 $\pm$ 0.083 & 0.142  \\ 
    &    &  0.544 $\pm$ 0.024 & & 2.523 $\pm$ 0.290 & &  \\   
 & I & 14.787 $\pm$ 0.426  & 19.403 $\pm$ 0.044 & 57.365 $\pm$
  4.728 & 18.994 $\pm$ 0.067 & 0.165  \\
   &   & 0.584 $\pm$ 0.017 & &  2.265 $\pm$ 0.187 & &  \\
\\ 
NGC 6814 & V & -   & - & - & - & - \\ 
  & & &  &  &  &  \\
 & R & 2.386 $\pm$ 0.196  & 17.862 $\pm$ 0.204 & 20.509 $\pm$
 1.732 & 18.955 $\pm$ 0.159 & 0.134  \\ 
   &     &  0.329 $\pm$ 0.027 & &  2.826 $\pm$ 0.239 & &  \\   
 & I & 2.908 $\pm$ 0.353  & 17.523 $\pm$ 0.296 & 20.116 $\pm$
  0.104 & 18.320 $\pm$ 0.010 & 0.157  \\
  &    & 0.401 $\pm$ 0.049 & &  2.771 $\pm$ 0.014 & &  \\
\\
Mrk 620 & V & 1.990 $\pm$ 0.696 & 18.825 $\pm$ 0.918 & 11.676 $\pm$
0.336 & 19.475 $\pm$ 0.062 & 0.191 \\ 
(NGC 2273) &  & 0.322 $\pm$ 0.112 &  & 1.893 $\pm$ 0.054  &  &  \\ 
 & R & 1.702 $\pm$ 0.165 & 17.829  $\pm$ 0.265 & 11.820 $\pm$
 0.458 & 19.102 $\pm$ 0.088 & 0.242 \\ 
   &     &  0.276 $\pm$ 0.027 & & 1.916 $\pm$ 0.074 & &  \\   
 & I & 1.677 $\pm$ 0.100  & 16.942 $\pm$ 0.156 & 11.787 $\pm$
  0.371 & 18.488 $\pm$ 0.069 & 0.304  \\
   &   & 0.272 $\pm$ 0.016 & &  1.911 $\pm$ 0.060 & &  \\
\\
Mrk 622  & V & 4.486 $\pm$ 0.113 & 20.807 $\pm$ 0.049 & 1.882 $\pm$
  0.105 & 16.110 $\pm$ 0.496 & 13.694 \\ 
  & & 2.754 $\pm$ 0.070 &  & 1.155 $\pm$ 0.064  &  &  \\ 
 & R & 3.787 $\pm$ 0.075 & 20.033  $\pm$ 0.040 & 2.340 $\pm$
 0.189 & 17.344 $\pm$ 0.584 & 15.330 \\ 
   &     &  2.324 $\pm$ 0.046 & & 1.436 $\pm$ 0.116 & &  \\  
 & I & 3.730 $\pm$ 0.085  & 19.496 $\pm$ 0.047 & 4.653 $\pm$
  0.871 & 19.869 $\pm$ 0.670 & 10.350 \\
   &   & 2.290 $\pm$ 0.052 & &  2.856 $\pm$ 0.535 & &  \\    
\hline
\end{tabular}

\end{table*}

\begin{table*}
\contcaption{}
\begin{tabular}{lcccccc}
\hline 
Galaxy & Filter & $r_{e}$ & $\mu_{e}$ & $r_{0}$ &
  $\mu_{0}$  & {\it B/D}\\
 & & (\arcsec )  & (mag/\arcsec$^{2}$) & (\arcsec ) & (mag/\arcsec$^{2}$) & \\
 & & (Kpc) &  & (Kpc) &  & \\
\hline
M 51 &  V & 8.989 $\pm$ 0.805 & 20.434 $\pm$ 0.151 & 17.811 $\pm$
  0.792 & 17.992 $\pm$ 0.066 & - \\ 
(NGC 5194)  & & 0.36 $\pm$ 0.033 &  & 0.727 $\pm$ 0.032  &  &  \\ 
 & R & 7.371 $\pm$ 0.510 & 19.786  $\pm$ 0.108 & 14.475 $\pm$
 0.582 & 17.337 $\pm$ 0.056 & -  \\ 
    &    &  0.301 $\pm$ 0.021 & & 0.591 $\pm$ 0.024 & &  \\   
 & I & 6.829 $\pm$ 0.600  & 19.031 $\pm$ 0.153 & 14.421 $\pm$
  0.474 & 16.714 $\pm$ 0.050 & -  \\
   &   & 0.279 $\pm$ 0.024 & &  0.588 $\pm$ 0.019 & &  \\ 
\\
NGC 1052 & V & 38.554 $\pm$ 0.964  & 21.918 $\pm$ 0.036 &  - & - & - \\ 
  & & 4.996 $\pm$ 0.125  &  &  &  &  \\
  & R & 35.321 $\pm$ 0.955  & 21.287 $\pm$ 0.040 & - & - & -  \\ 
  & & 4.577 $\pm$ 0.124  &  &  &  &  \\   
 & I & 28.426  $\pm$ 1.042  & 20.285 $\pm$ 0.055 &  - & - & - \\ 
  & & 3.683 $\pm$ 0.135 & & &  &  \\ 
\\
NGC 2841 & V & 33.019 $\pm$ 0.735 & 21.513 $\pm$ 0.030 & 54.806 $\pm$
  12.437 & 21.344 $\pm$ 0.151 & 1.121 \\ 
  & & 1.857 $\pm$ 0.041 &  & 3.082 $\pm$ 0.699  &  &  \\
  & R & 29.150 $\pm$ 0.456 & 20.870  $\pm$ 0.022 & 49.690 $\pm$
 7.206 & 20.726 $\pm$ 0.110 & 1.088 \\ 
     &   &  1.639 $\pm$ 0.026 & & 2.794 $\pm$ 0.405 & &  \\  
 & I & 26.658 $\pm$ 0.371  & 20.052 $\pm$ 0.020 & 54.313 $\pm$
  2.231 & 19.952 $\pm$ 0.026 & 0.793  \\
   &   & 1.499 $\pm$ 0.021 & &  3.054 $\pm$ 0.125 & &  \\ 
\\
NGC 3998 & V & 10.363 $\pm$ 0.267  & 19.365 $\pm$ 0.048 & 17.987 $\pm$
  1.064 & 20.173 $\pm$ 0.139 & 5.011 \\ 
  & & 0.973 $\pm$ 0.025 & & 1.689 $\pm$ 0.100 &  &  \\
  & R & 9.627 $\pm$ 0.188  & 18.797 $\pm$ 0.036 & 17.120 $\pm$
 1.099 & 19.746 $\pm$ 0.157 & 5.734  \\ 
    &    &  0.904 $\pm$ 0.018 & &  1.607 $\pm$ 0.103 & &  \\
 & I & 9.855 $\pm$ 0.178  & 18.098 $\pm$ 0.033 & 17.404 $\pm$
  1.355 & 19.079 $\pm$ 0.187 & 5.880 \\
   &   & 0.925 $\pm$ 0.017 & &  1.634 $\pm$ 0.127 & &  \\
\\
NGC 5077 & V & 16.889 $\pm$ 0.598  & 21.201 $\pm$ 0.059 &  - & - & - \\ 
  & & 4.216 $\pm$ 0.149  &  &  &  &  \\ 
  & R & 15.385 $\pm$ 0.362  & 20.468 $\pm$ 0.040 & - & - & -  \\ 
  & & 3.841 $\pm$ 0.090  &  &  &  &  \\   
  & I & 15.355 $\pm$ 0.466  & 20.095 $\pm$ 0.051 &  - & - & - \\ 
  &  & 3.883 $\pm$ 0.116 & & &  &  \\ 
\\
NGC 6340 & V & 9.251 $\pm$ 0.326 & 20.486 $\pm$ 0.066 & 22.669 $\pm$
  0.325 & 20.150 $\pm$ 0.029 & 0.441 \\ 
  & & 0.976 $\pm$ 0.034 &  & 2.392 $\pm$ 0.034  &  &  \\ 
  & R & 8.488 $\pm$ 0.238 & 19.926 $\pm$ 0.054 & 22.088 $\pm$
 0.269 & 19.756 $\pm$ 0.026 & 0.455  \\ 
    &    &  0.896 $\pm$ 0.025 & & 2.331 $\pm$ 0.028 & &  \\  
  & I & 8.610 $\pm$ 0.190  & 19.227 $\pm$ 0.041 & 23.031 $\pm$
  0.340 & 19.090 $\pm$ 0.030 & 0.445 \\
  &   & 0.908 $\pm$ 0.020 & &  2.430 $\pm$ 0.036 & &  \\
\\
NGC 6384 & V & 9.788 $\pm$ 0.352 & 20.535 $\pm$ 0.072 & 27.015 $\pm$
  1.979 & 20.619 $\pm$ 0.108 & 0.512 \\ 
  & & 1.435 $\pm$ 0.052 &  & 3.960 $\pm$ 0.290  &  &  \\
  & R & 2.981 $\pm$ 0.337 & 17.768 $\pm$ 0.307 & 22.931 $\pm$
 1.645 & 19.671 $\pm$ 0.143 &  0.352 \\ 
    &    &  0.437 $\pm$ 0.049 & & 3.361 $\pm$ 0.241 & &  \\  
 & I & 4.957 $\pm$ 0.245  & 18.252 $\pm$ 0.118 & 21.506 $\pm$
  1.229 & 19.138 $\pm$ 0.106 & 0.434 \\
   &   & 0.727 $\pm$ 0.036 & &  3.153 $\pm$ 0.180 & &  \\
\\
NGC 7177 & V & 12.885 $\pm$ 1.254 & 20.913 $\pm$ 0.170 & 10.897 $\pm$
  0.210 & 19.777 $\pm$ 0.057 & 1.772 \\ 
  & & 1.305 $\pm$ 0.127 &  & 1.104 $\pm$ 0.021  &  &  \\ 
  & R & 11.598 $\pm$ 1.018 & 20.253  $\pm$ 0.158 & 11.588 $\pm$
 0.318 & 19.580 $\pm$ 0.074 &  1.946 \\ 
   &    &  1.175 $\pm$ 0.103 & & 1.175 $\pm$ 0.032 & &  \\  
  & I & 6.547 $\pm$ 0.479  & 18.575 $\pm$ 0.153 & 11.169 $\pm$
  0.325 & 18.541 $\pm$ 0.083 & 1.202  \\
  &    & 0.663 $\pm$ 0.049 & &  1.131 $\pm$ 0.033 & &  \\
\\
NGC 7217 & V & 20.358 $\pm$ 1.611 & 20.885 $\pm$ 0.121 & 19.623 $\pm$
  0.570 & 19.220 $\pm$ 0.059 & 0.838 \\ 
  & & 1.697 $\pm$ 0.134 &  & 1.636 $\pm$ 0.047  &  &  \\
  & R & 30.241 $\pm$ 2.622 & 20.878  $\pm$ 0.120 & 19.863 $\pm$
 0.648 & 19.029 $\pm$ 0.065 & 1.525 \\ 
    &    &  2.521 $\pm$ 0.219 & & 1.656 $\pm$ 0.054 & &  \\  
  & I & 38.220 $\pm$ 2.514  & 20.487 $\pm$ 0.086 & 17.819 $\pm$
  0.636 & 18.481 $\pm$ 0.079 & 2.617  \\
    &  & 3.187 $\pm$ 0.210 & &  1.486 $\pm$ 0.053 & &  \\
\\
\hline
\end{tabular}

\end{table*}

\begin{table*}
\contcaption{}
\begin{tabular}{lcccccc}
\hline 
Galaxy & Filter & $r_{e}$ & $\mu_{e}$ & $r_{0}$ &
  $\mu_{0}$  & {\it B/D}\\
 & & (\arcsec)  & (mag/\arcsec$^{2}$) & (\arcsec) & (mag/\arcsec$^{2}$) & \\
 & & (Kpc) &  & (Kpc) &  & \\
\hline
NGC 2146 & V & -  & - & - & - & - \\ 
  & & &  & &  &  \\
  & R & - & - & - & - & -  \\ 
     &   & & & & &  \\  
  & I & 8.444 $\pm$ 1.322 & 20.351 $\pm$ 0.278 & 20.136 $\pm$
  1.055 & 19.173 $\pm$ 0.092 & 0.214 \\
    &  & 0.665 $\pm$ 0.104 & &  1.585 $\pm$ 0.083 & &  \\
\\
NGC 3310 & V & 1.950 $\pm$ 0.049 & 18.431 $\pm$ 0.045 & 12.808 $\pm$
  0.637 & 19.382 $\pm$ 0.185 & 0.201 \\ 
  & & 0.168 $\pm$ 0.004 &  & 1.106 $\pm$ 0.055  &  &  \\ 
  & R & 2.042 $\pm$ 0.211 & 18.058  $\pm$ 0.187 & 13.346 $\pm$
 0.975 & 19.245 $\pm$ 0.231 & 0.252  \\ 
     &   &  0.176 $\pm$ 0.018 & & 1.153 $\pm$ 0.084 & &  \\  
 & I & 2.563 $\pm$ 0.305  & 17.932 $\pm$ 0.205 & 12.357 $\pm$
  0.567 & 18.679 $\pm$ 0.175 & 0.309 \\
   &   & 0.221 $\pm$ 0.026 & &  1.067 $\pm$ 0.049 & &  \\
\\
NGC 6217 & V & 0.708 $\pm$ 0.019 & 16.528 $\pm$ 0.084 & 21.276 $\pm$
  1.176 & 20.137 $\pm$ 0.058 & 0.111 \\ 
  & & 0.085 $\pm$ 0.002 &  & 2.554 $\pm$ 0.141  &  &  \\
  & R & 1.425 $\pm$ 0.057  & 17.818 $\pm$ 0.105 & 20.557 $\pm$
 1.035 & 19.805 $\pm$ 0.054 & 0.108 \\ 
   &  &  0.171 $\pm$ 0.007 & & 2.468 $\pm$ 0.124 & &  \\  
 & I & 3.192 $\pm$ 0.109  & 18.974 $\pm$ 0.075 & 20.353 $\pm$
  0.770 & 19.276 $\pm$ 0.041 & 0.117 \\
    &  & 0.383 $\pm$ 0.013 & &  2.443 $\pm$ 0.092 & &  \\
\\ 
NGC 1023 &  V & 21.680 $\pm$ 0.872 & 20.627 $\pm$ 0.058 & 29.703 $\pm$
  1.879 & 20.506 $\pm$ 0.062 & 1.721 \\ 
 &   & 1.217 $\pm$ 0.049 &  & 1.667 $\pm$ 0.105  &  &  \\ 
 & R & 19.607 $\pm$ 0.673 & 19.967 $\pm$ 0.051 & 27.754 $\pm$
 1.662 & 19.979 $\pm$ 0.064 & 1.822  \\ 
    &    &  1.101 $\pm$ 0.038 & & 1.558 $\pm$ 0.093 & &  \\  
  & I & 19.731 $\pm$ 0.253 & 19.333 $\pm$ 0.019 & 25.952 $\pm$
  0.903 & 19.354 $\pm$ 0.040 & 2.127 \\ 
  & & 1.108 $\pm$ 0.014 &  & 1.457 $\pm$ 0.051  &  &  \\
\hline
\end{tabular}

\end{table*}

\begin{figure*}
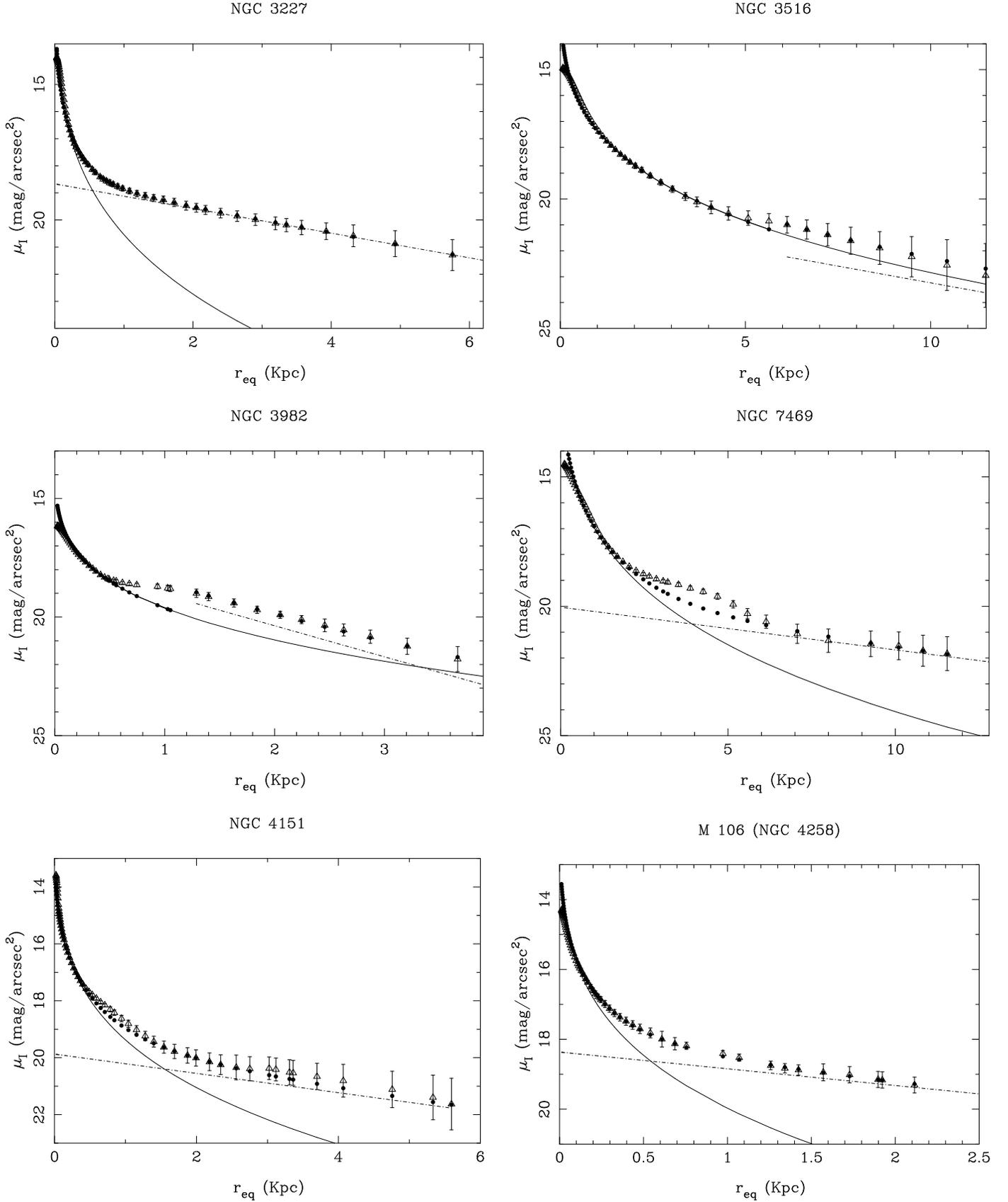

\hbox{
\hspace{-3mm}
\psfig{figure=parn3227i.ps,width=90mm,angle=270}
\hspace{3mm}
\psfig{figure=parn3516i.ps,width=90mm,angle=270}
}
\vspace{12pt}
\hbox{
\hspace{-3mm}
\psfig{figure=parn3982i.ps,width=90mm,angle=270}
\hspace{3mm}
\psfig{figure=parn7469i.ps,width=90mm,angle=270}
}
\vspace{12pt}
\hbox{
\hspace{-3mm}
\psfig{figure=parn4151i.ps,width=90mm,angle=270}
\hspace{3mm}
\psfig{figure=parn4258i.ps,width=90mm,angle=270}
}
\caption{$I$ band parameter decomposition. $\triangle$ symbols 
  represent the calibrated surface
  brightness profile, the solid line shows the bulge model, the
  dash-dotted line represents the disc model and $\bullet$ symbols
  show the sum of bulge and disc models.}
\label{fig:parametros1}
\end{figure*}

\begin{figure*}
\hbox{
\hspace{-3mm}
\psfig{figure=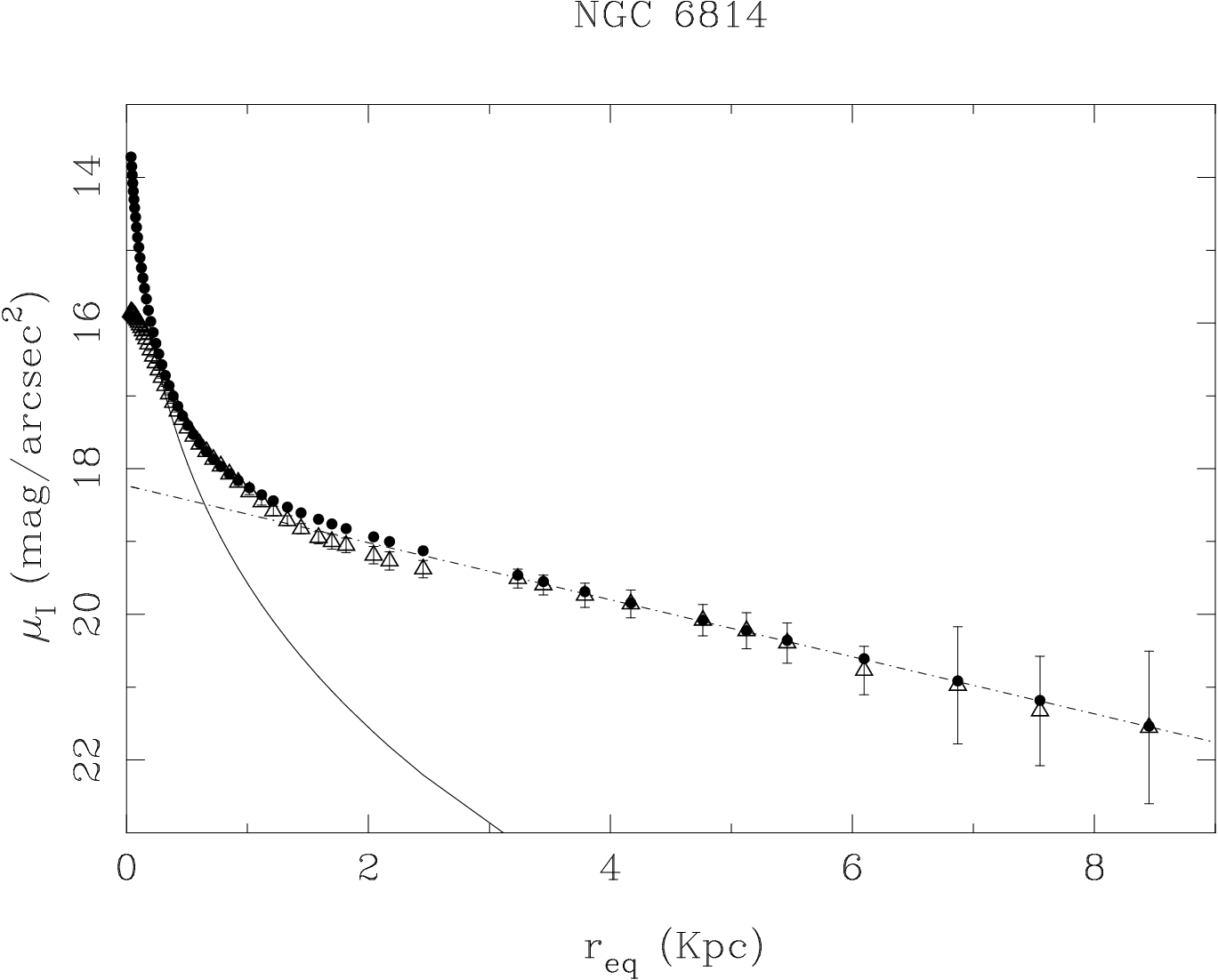,width=90mm,angle=270}
\hspace{3mm}
\psfig{figure=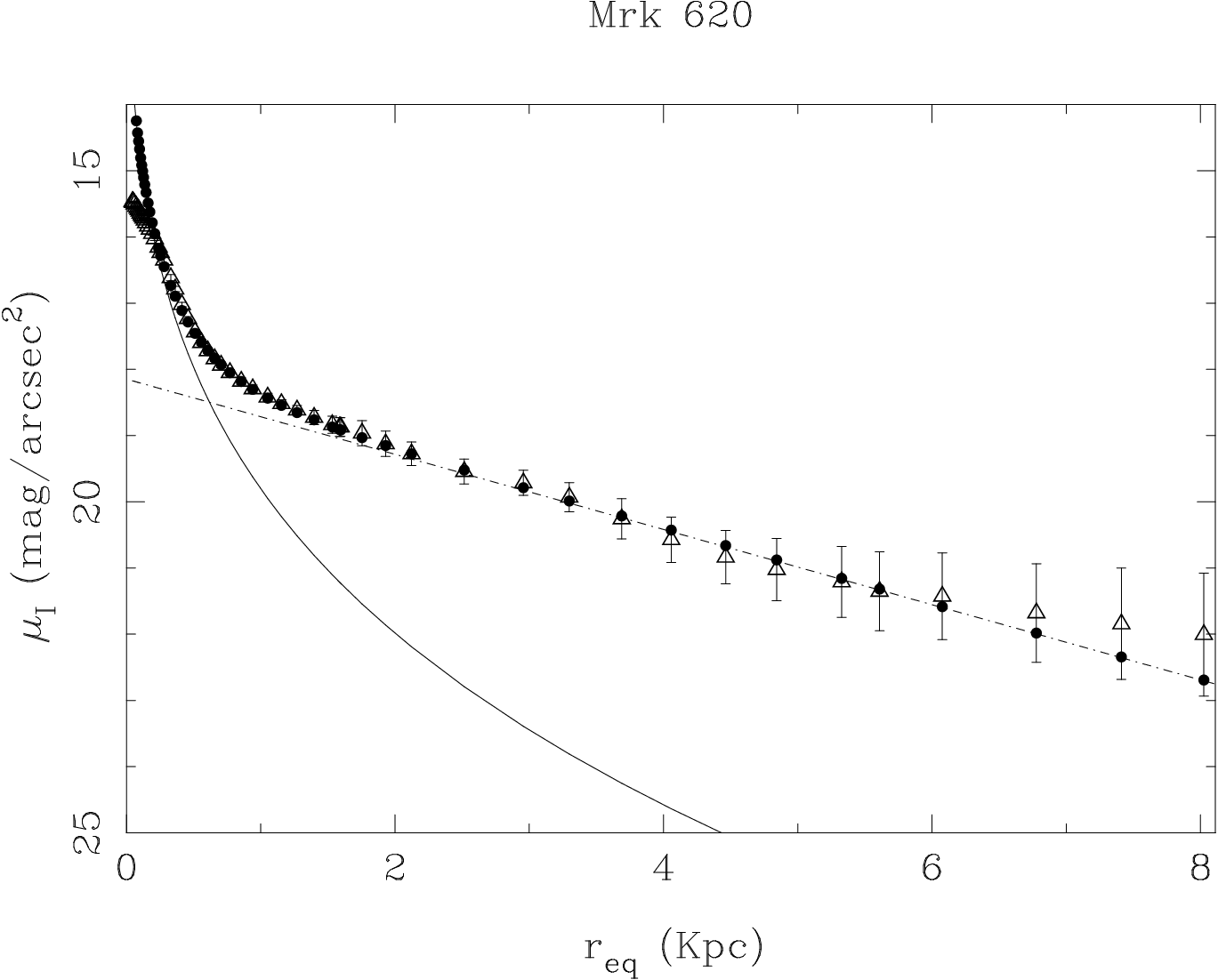,width=90mm,angle=270}
}
\vspace{12pt}
\hbox{
\hspace{-3mm}
\psfig{figure=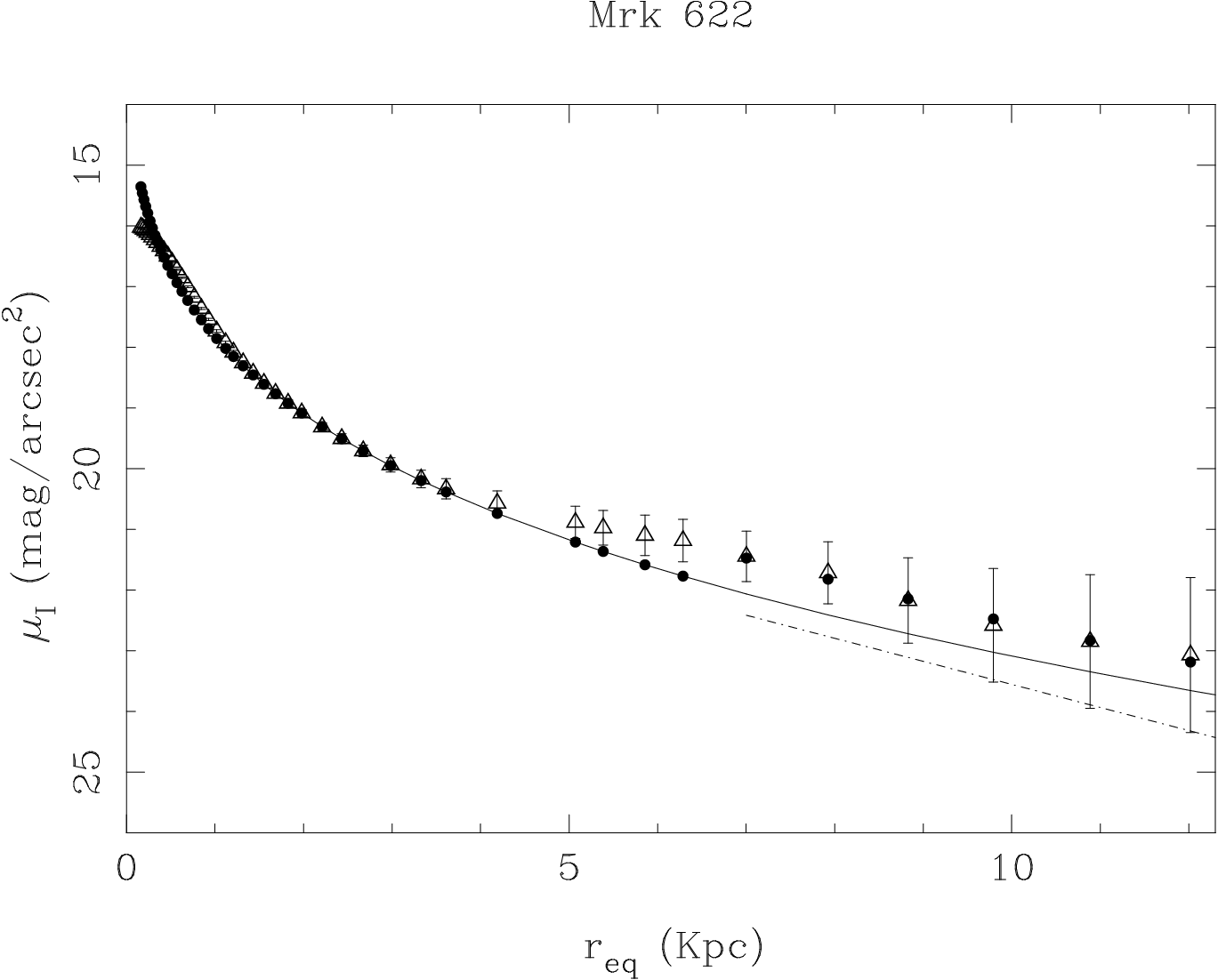,width=90mm,angle=270}
\hspace{3mm}
\psfig{figure=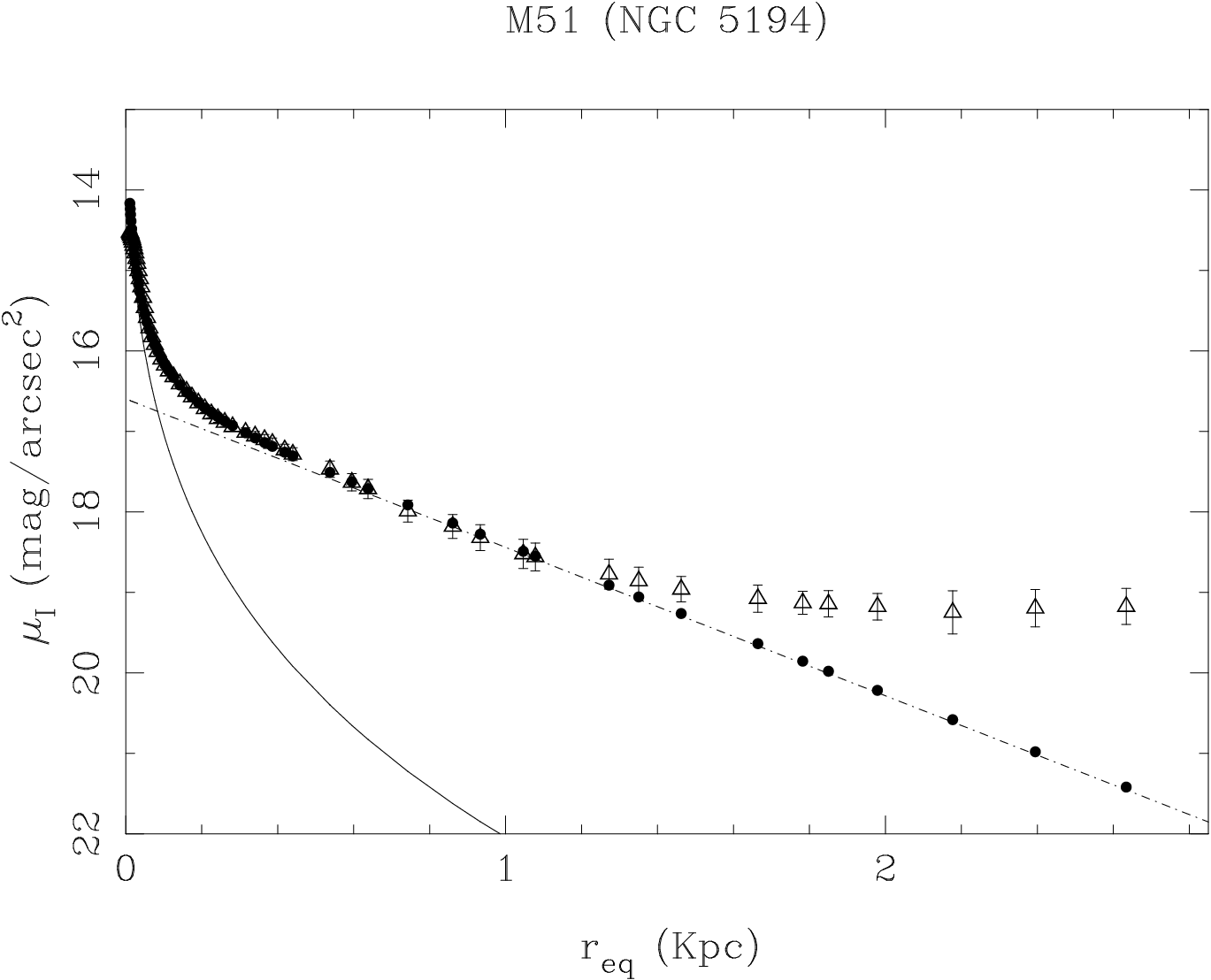,width=90mm,angle=270}
}
\vspace{12pt}
\hbox{
\hspace{-3mm}
\psfig{figure=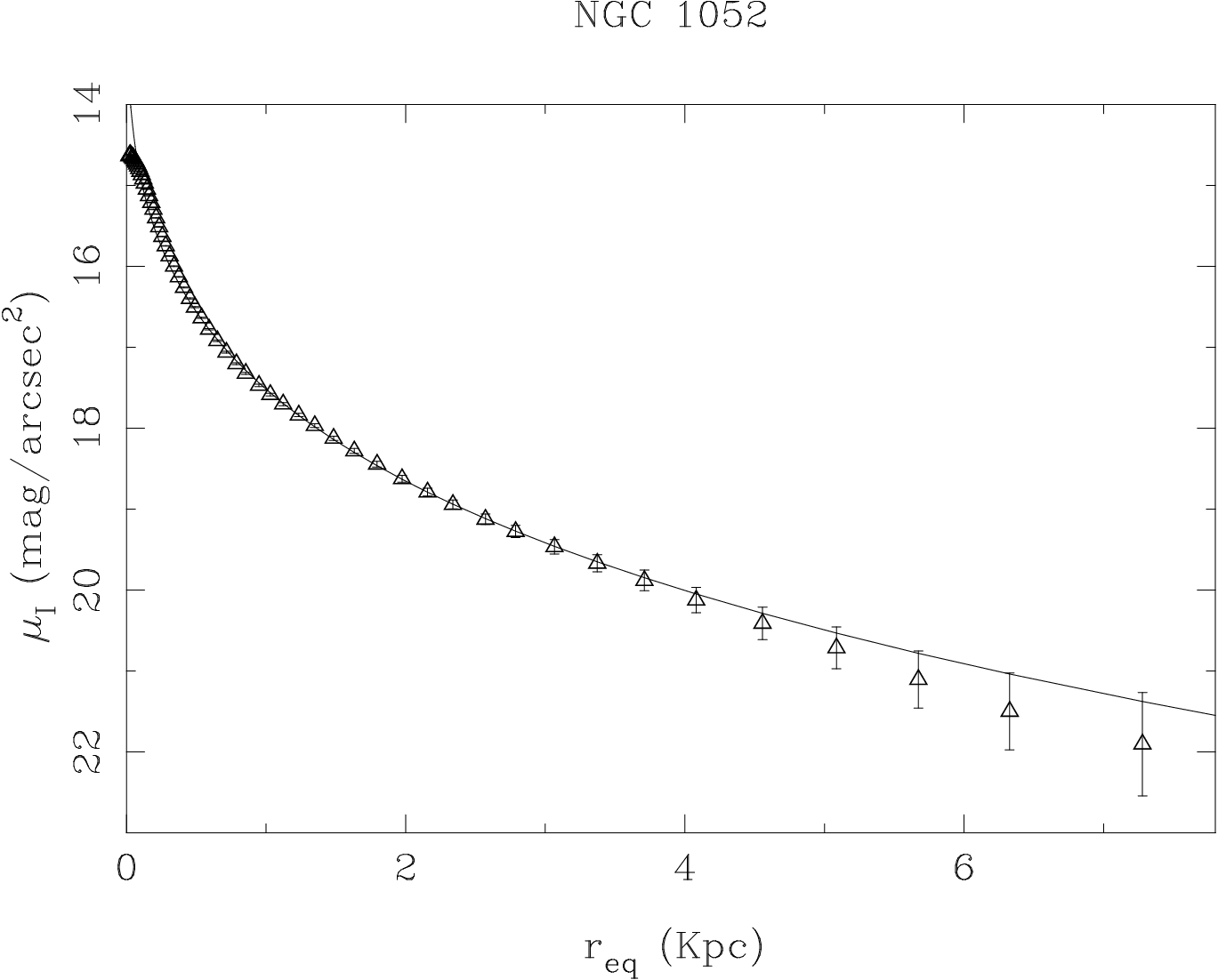,width=90mm,angle=270}
\hspace{3mm}
\psfig{figure=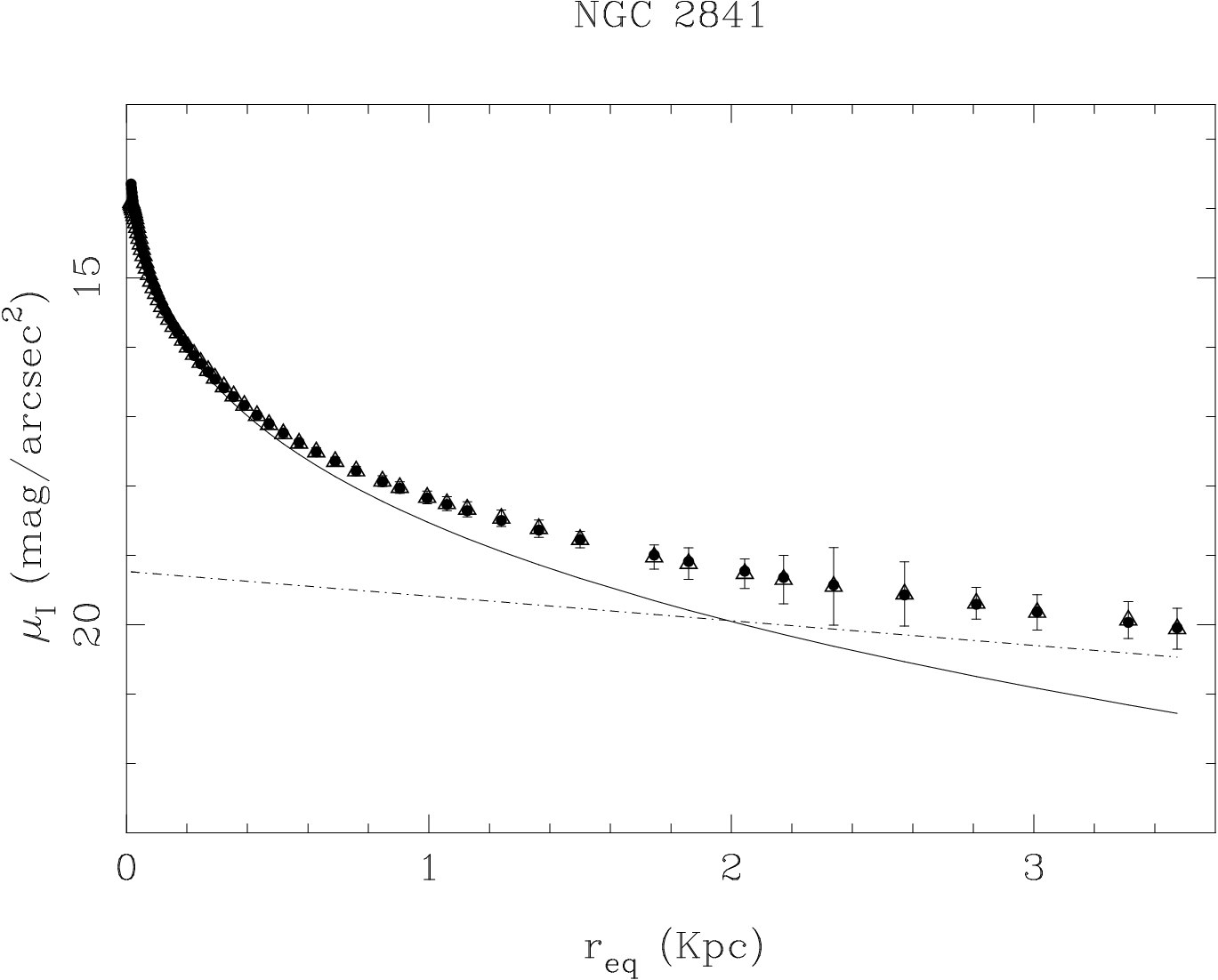,width=90mm,angle=270}
}
\contcaption{}
\end{figure*}

\begin{figure*}
\hbox{
\hspace{-3mm}
\psfig{figure=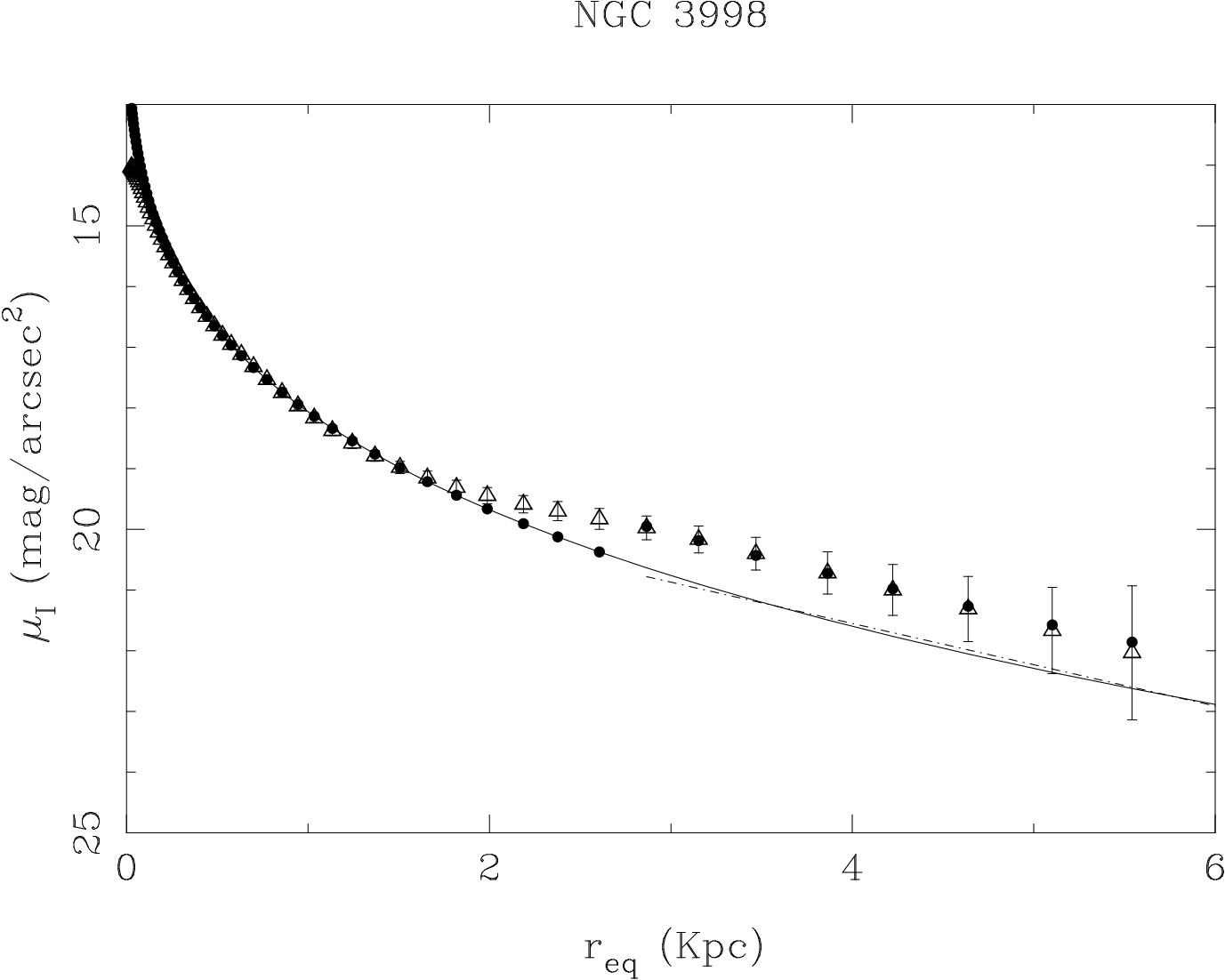,width=90mm,angle=270}
\hspace{3mm}
\psfig{figure=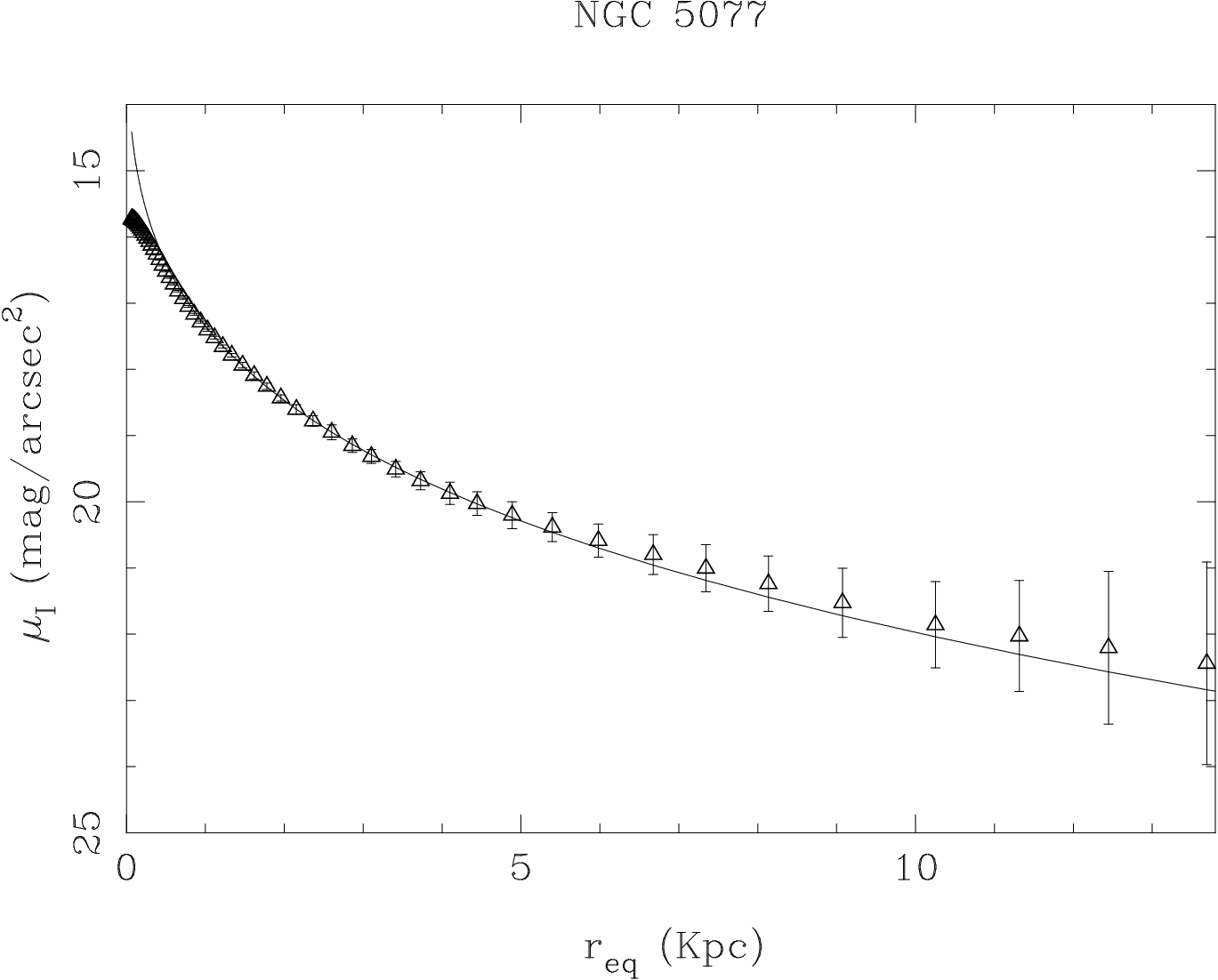,width=90mm,angle=270}
}
\vspace{12pt}
\hbox{
\hspace{-3mm}
\psfig{figure=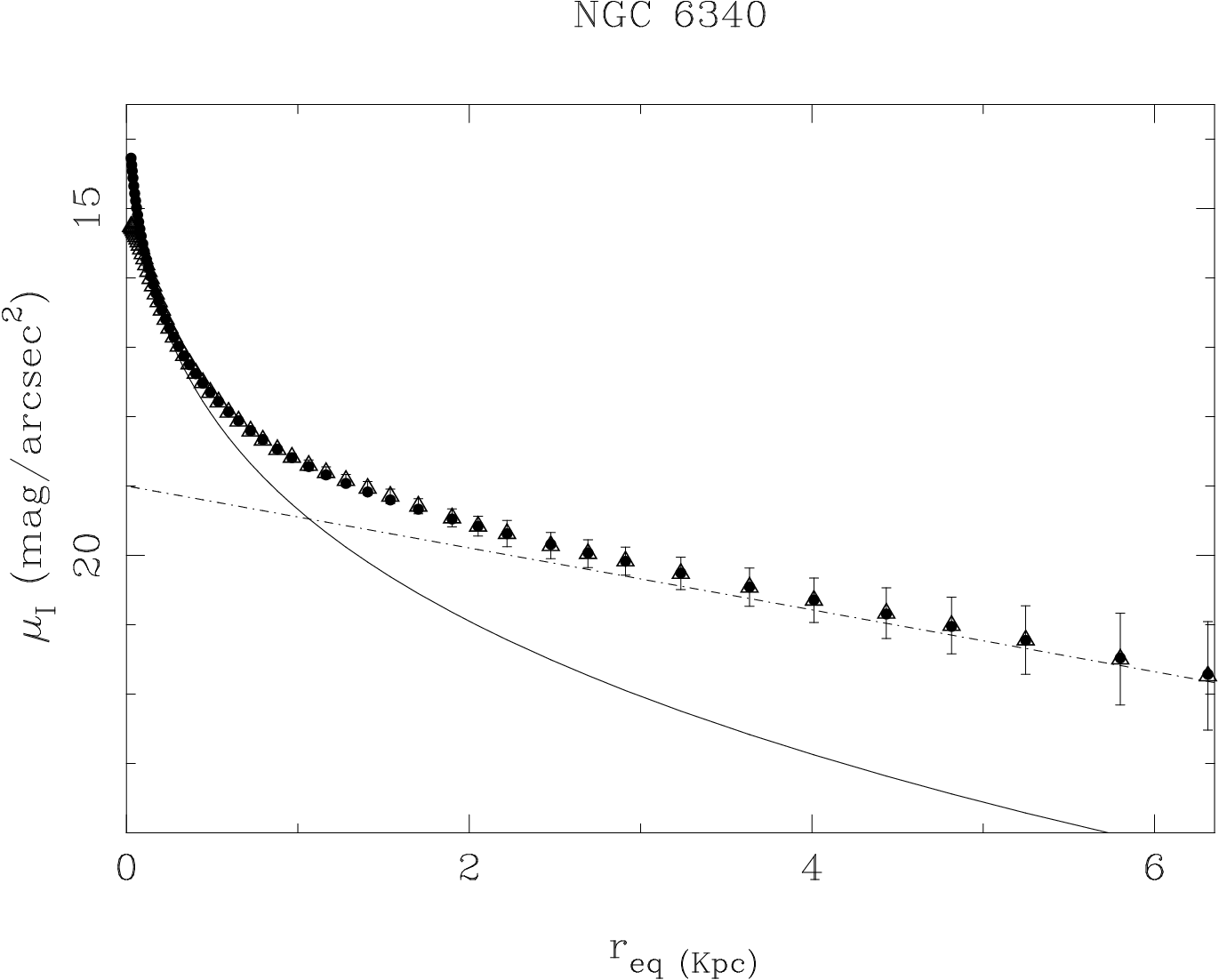,width=90mm,angle=270}
\hspace{3mm}
\psfig{figure=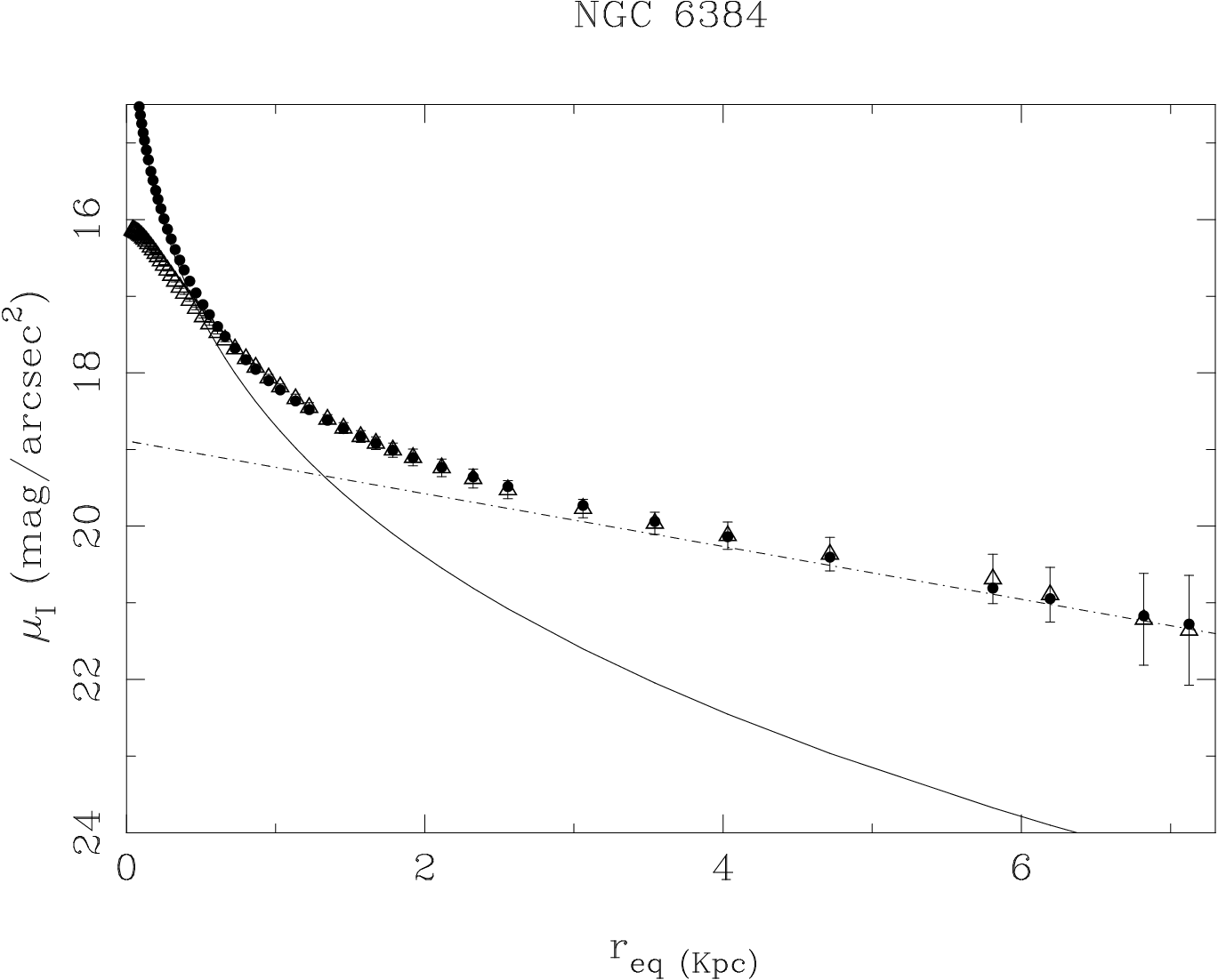,width=90mm,angle=270}
}
\vspace{12pt}
\hbox{
\hspace{-3mm}
\psfig{figure=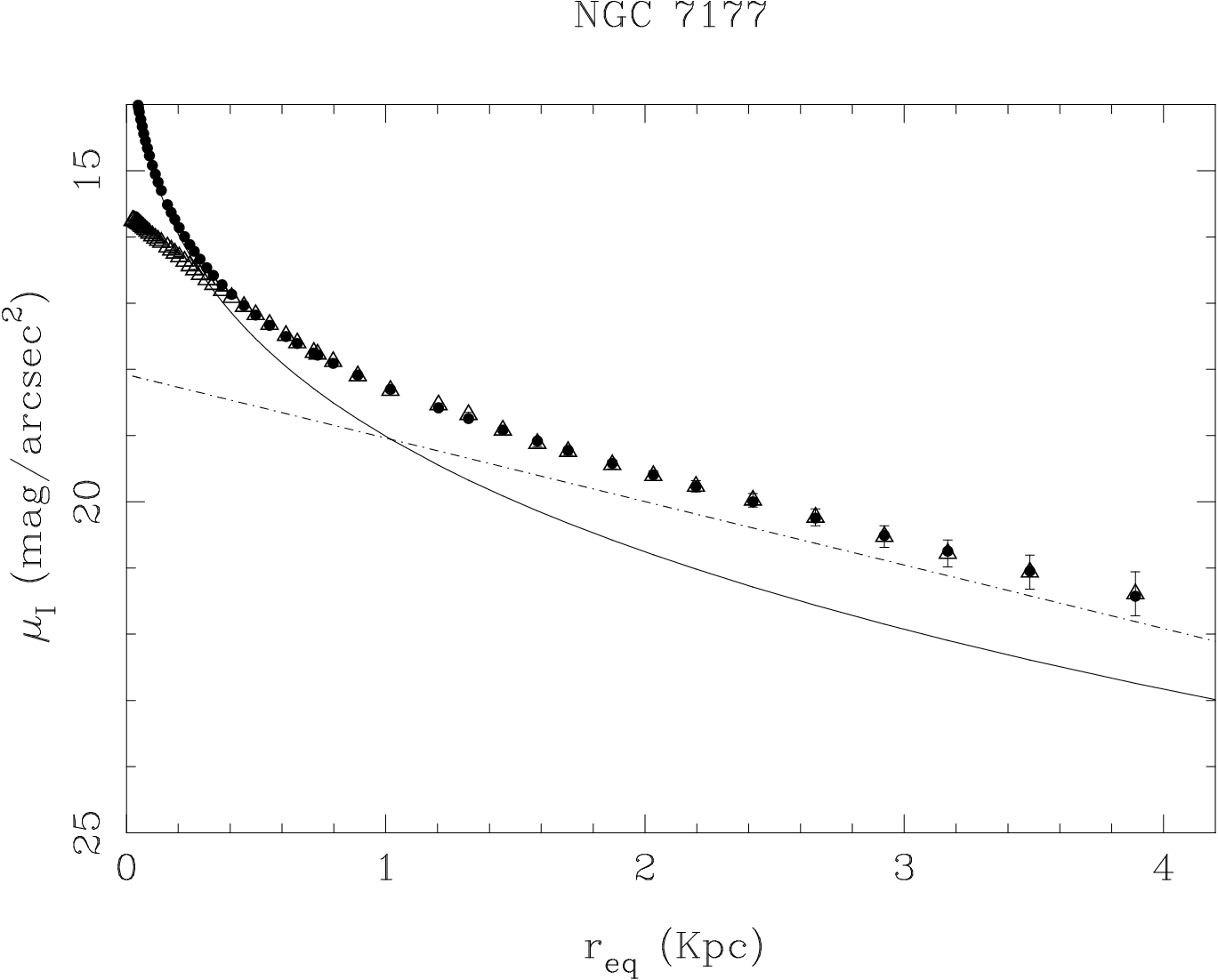,width=90mm,angle=270}
\hspace{3mm}
\psfig{figure=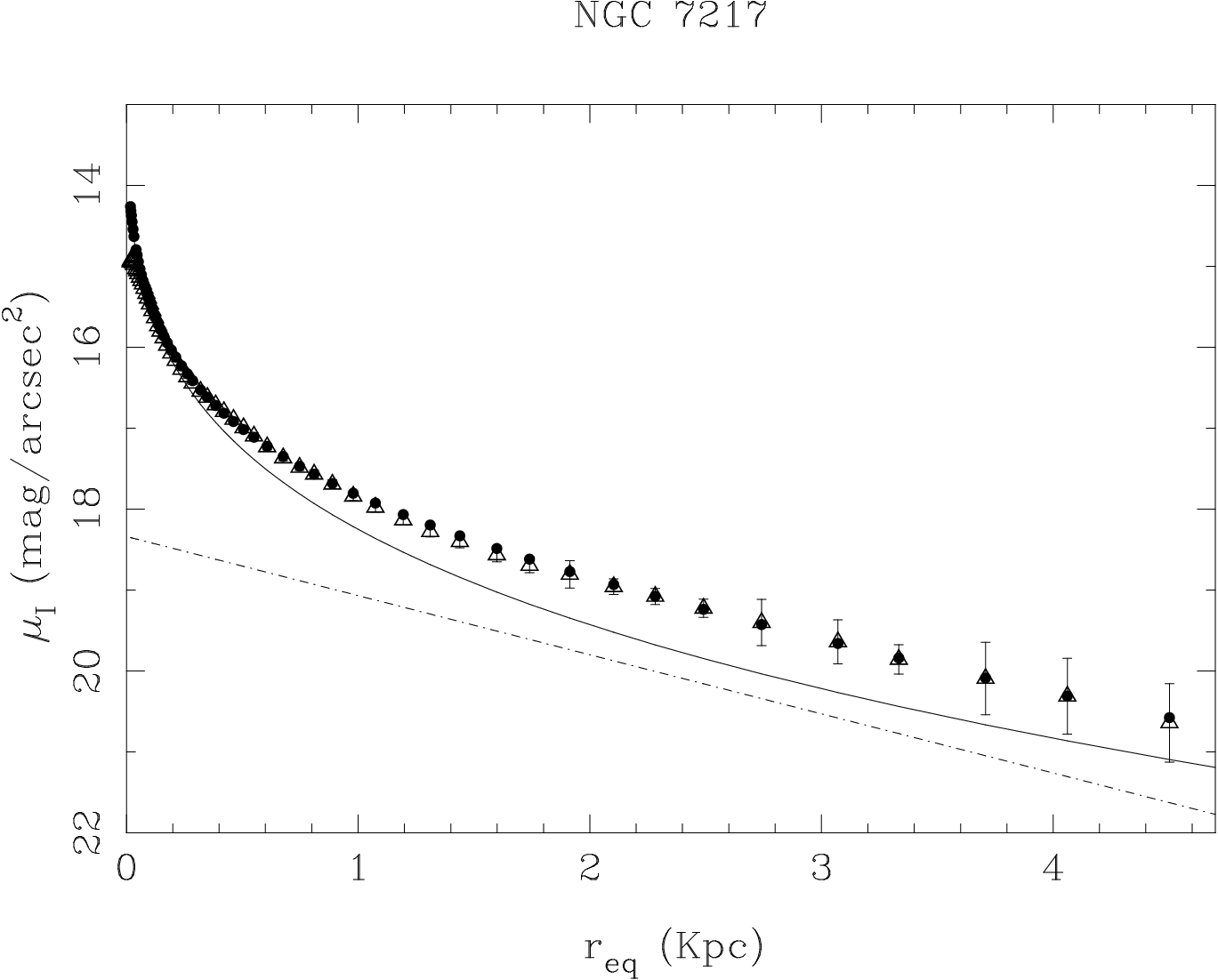,width=90mm,angle=270}
}
\contcaption{}
\end{figure*}

\begin{figure*}
\hbox{
\hspace{-3mm}
\psfig{figure=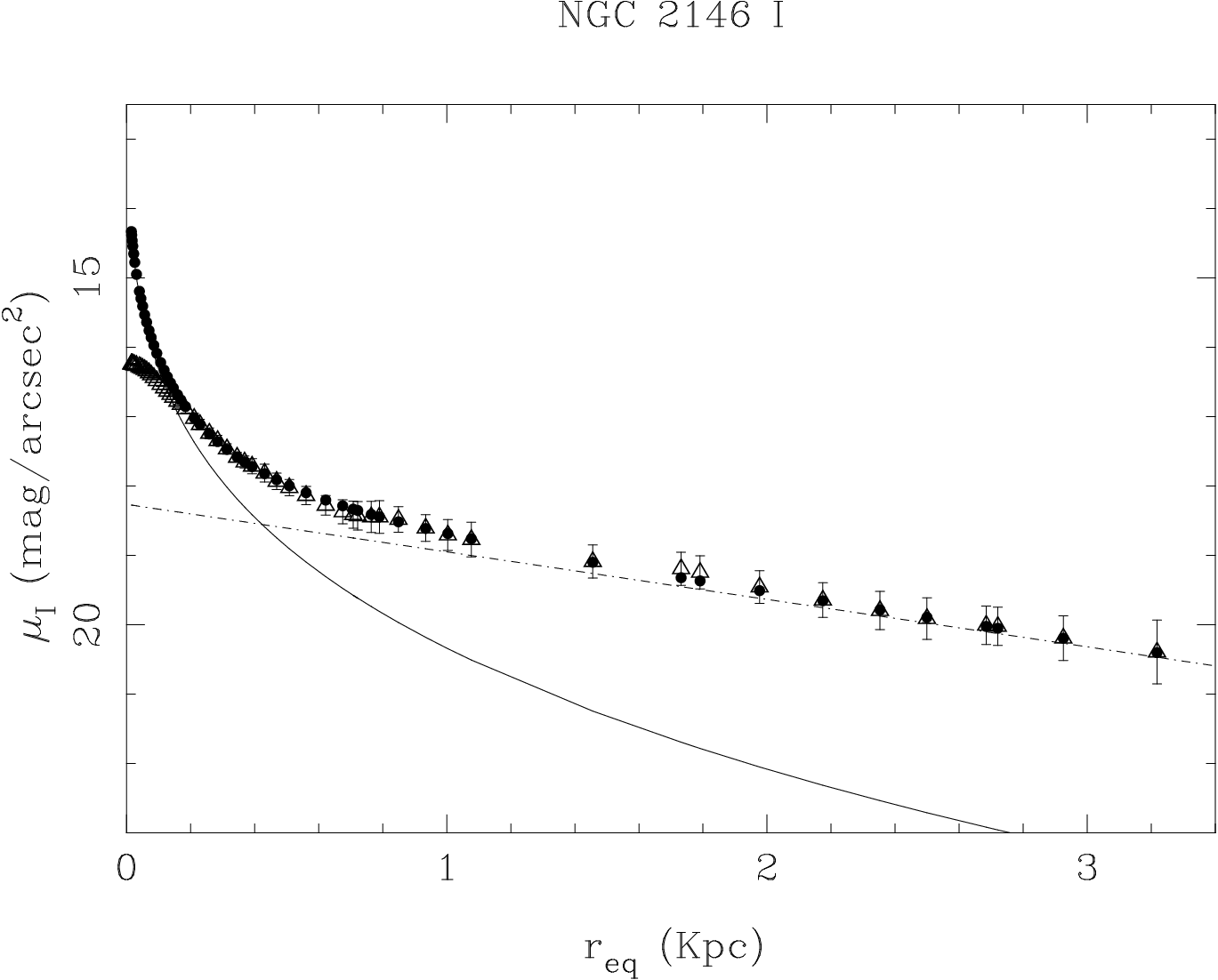,width=90mm,angle=270}
\hspace{3mm}
\psfig{figure=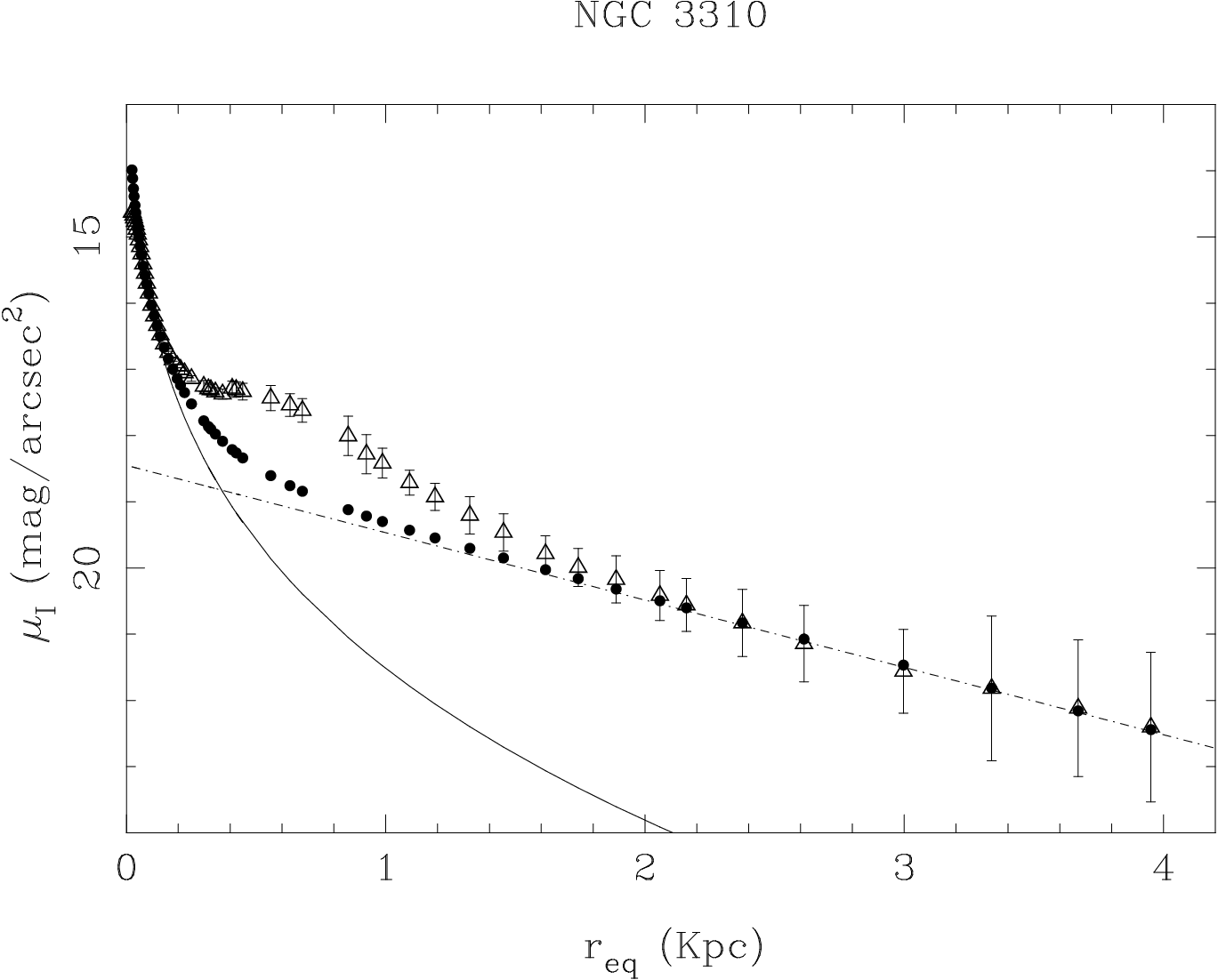,width=90mm,angle=270}
}
\vspace{12pt}
\hbox{
\hspace{-3mm}
\psfig{figure=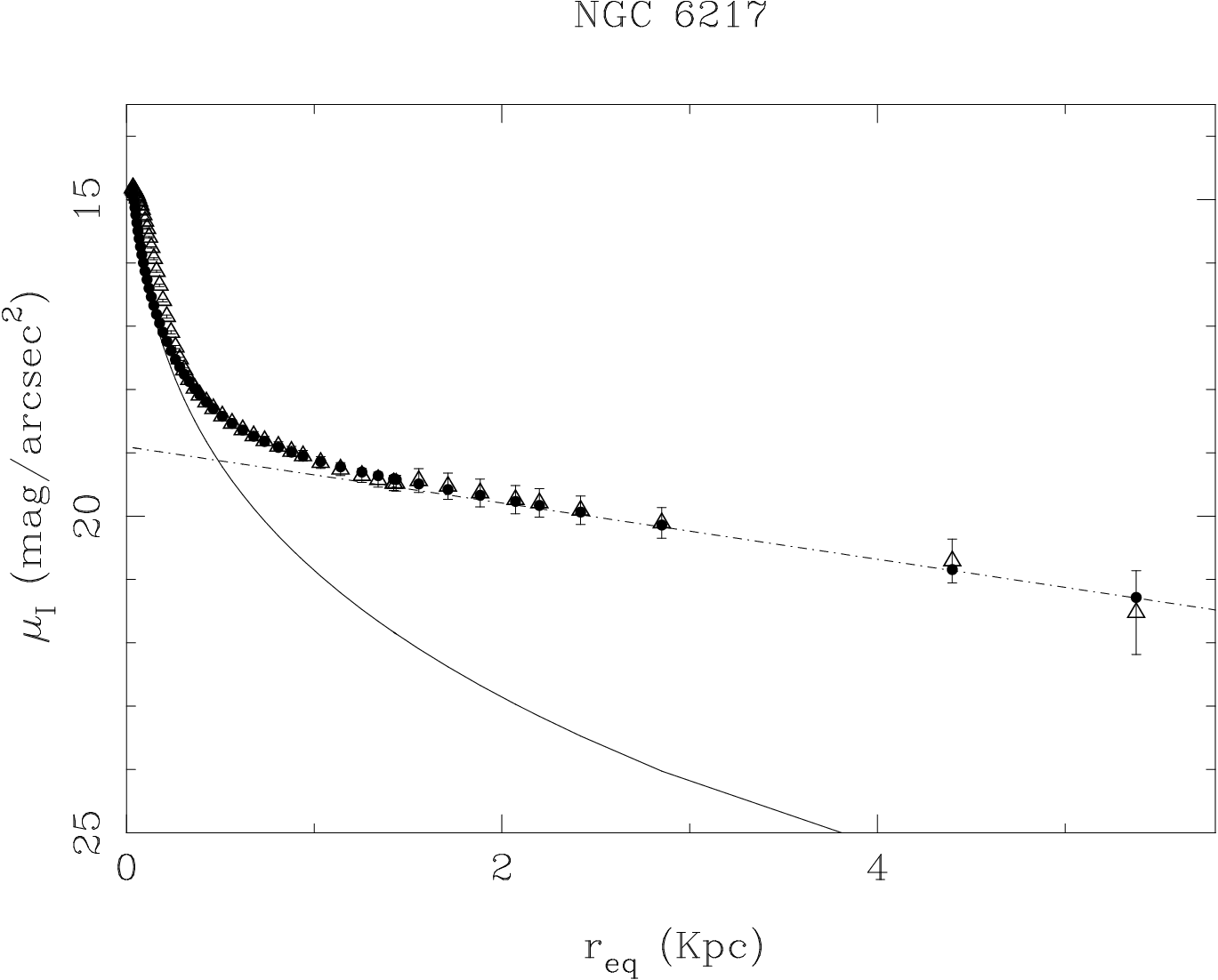,width=90mm,angle=270}
\hspace{3mm}
\psfig{figure=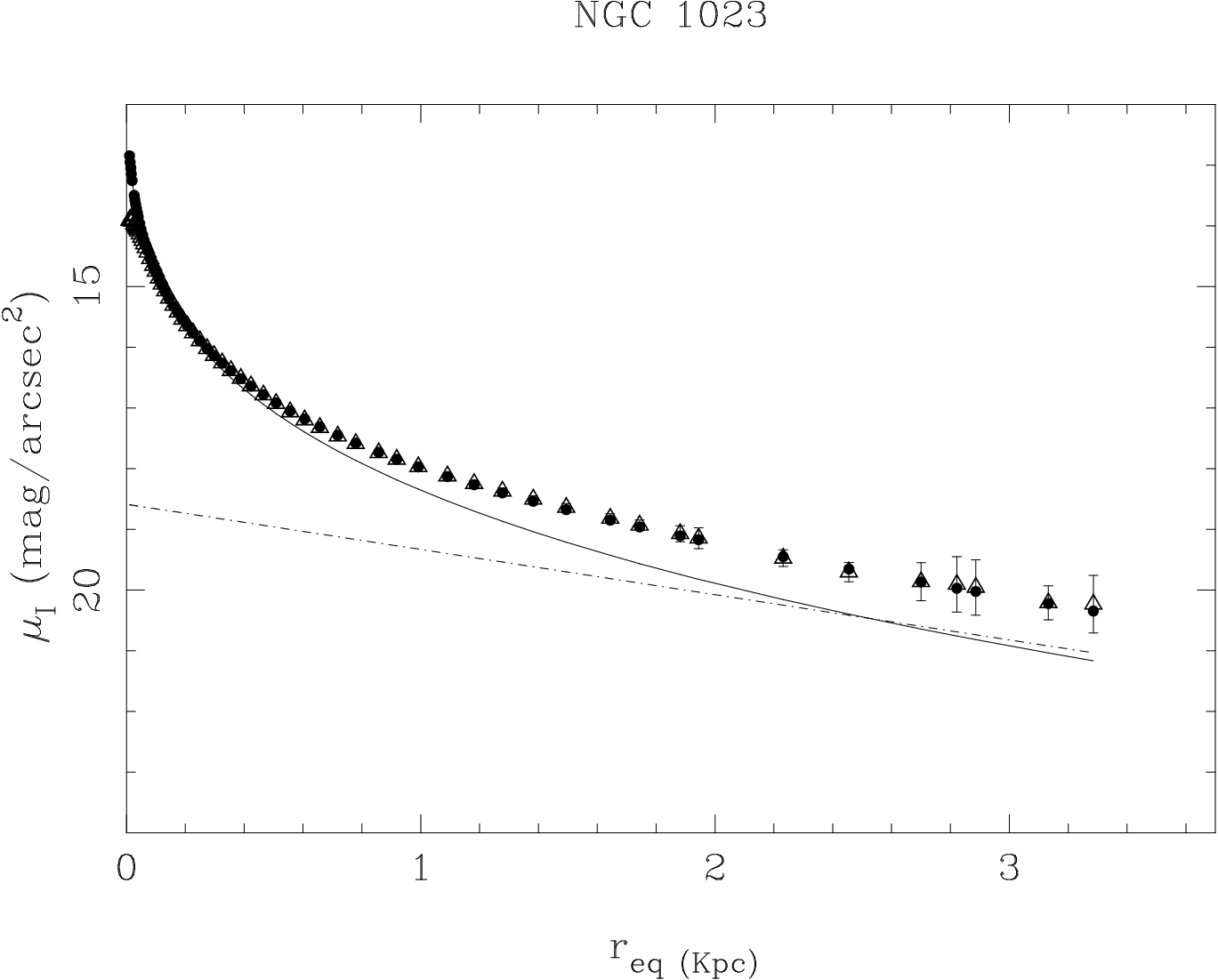,width=90mm,angle=270}
}
\contcaption{}
\end{figure*}

\subsection{Notes on individual objects}
\label{notes}
\subsubsection{Seyfert~1.x galaxies}
\begin{itemize}

\item NGC~3516: we have fitted to this SB0$^0$ galaxy a Freeman
  type II disc\footnote{As explained before, in all calculations
    throughout this paper, we 
  have assumed a simplified type II disc with a sharp cutoff at
  $r_{cut}$.} with $r_{cut} = 5.7$~Kpc (25\arcsec)

\item  NGC~7469: a large excess for $ 2 \lsim r \lsim 6$~Kpc is
  observed in all the  surface brightness profiles. M\'arquez \& Moles
  \shortcite{AJ108:Marquez}  atribute this feature to a lens
  structure, although they did not perform an accurate numerical fitting.  

\item  NGC~4151: as already commented in Paper~I, this spiral SABab
  presents a large-scale bar that affects the surface brightness
  profile from \mbox{$ r \sim 2.5 $~Kpc} outwards. Therefore, the disc profile
  is somewhat uncertain.

\item  NGC~6814: this SABbc galaxy presents a quite compact
  bulge. This fact, along with the relatively poor seeing (\mbox{$\sim$
  2.3\arcsec FWHM}) prevented us from obtaining $V$ band morphological parameters.
M\' arquez et al.  \shortcite{MarquezI} observed this galaxy in the $J$ and
$K'$ bands, finding bar parameters ($PA$, $\epsilon$) very similar to ours 
(see Paper~I). 
Nevertheless, the bulge and disc morphological parameters derived in the 
NIR are quite different to our optical parameters (though comparable):
the bulge and disc equivalent radius lie between 9.2~arcsec
 and 62~arcsec in the $J$ band 
and 5.5~arcsec and 40~arcsec in the $K'$ band while our $I$ band 
bulge and disc equivalent radius are equal to 2.9~arcsec and 33.8~arcsec, 
respectively. 

\item NGC~1068: an abrupt fall in the surface brightness profile (see
  Paper~I) is observed at  $r \approx  2.2$~Kpc, leading to an
  ``inner'' and an ``outer'' disc. Alternatively, this feature in the surface
brightness profile can be atributed to the existence of a circumnuclear
star-forming ring whose intensity peaks around 1.5~Kpc (D\'{\i}az et al. 2000). 
This fact, along with the 
  poor seeing conditions during the observations ($\sim$ 2.2\arcsec FWHM)
prevented us from performing an acceptable parameter decomposition.

\item NGC~3982: this is the latest type (SABb) galaxy that presents a
  type II disc. We have applied a fit with $ r_{cut} = 12\arcsec$.

\end{itemize}

\subsubsection{Seyfert~2 galaxies}

\begin{itemize}

\item NGC~513: in Paper~I we classified this galaxy as
  SABa. We failed to perform a simple bulge+disc decomposition, but perhaps 
  this galaxy owns a type II disc or additional components like a
  ring or lens.

\item  Mrk~622: according to Paper~I, we classified this galaxy as
  SB0. We have fitted a type II disc with $ r_{cut} = 11''$.

\item M~51: the overall disc does not obey an exponential law. For 
$ r > 40''$ the galaxy shows a flat profile, while for smaller radii
the surface brightness distribution can be fitted to an exponential
profile with small scalelength. This behaviour has been also reported
by Boroson \shortcite{Boroson}, who finds a flat disc for $ r >
52''$. The shape of the surface brightness profile can be explained by
the presence of a nearby companion galaxy, NGC~5195.

\item NGC~5347: the outstanding bar present in this SBab galaxy made
  impossible to perform a bulge+disc parameter decomposition. 
The $K'$ image from M\' arquez et al. \shortcite{MarquezI} only shows a
large bar also observed in our optical images. Their NIR bar parameters 
($PA$, $\epsilon$) match our optical parameters (Paper~I). 
They are able to derive bulge and disc morphological parameters:
the bulge and disc equivalent radius lie between 7.9~arcsec
 and 23~arcsec, respectively, in the $J$ band 
and 2.2~arcsec and 19~arcsec in the $K'$ band. 

\item NGC~7479: as in the previous case, this SBc has a very
  strong bar that dominates the surface profile. Therefore, it was not
  possible to perform an acceptable parameter decomposition.

\end{itemize}

\subsubsection{LINERs}

\begin{itemize}

\item  NGC~3998:  we have fitted to this SA0$^0$ galaxy a Freeman
  type II disc with $r_{cut} = 2.75$~Kpc (30\arcsec).

\item  NGC~5077: this E3-4 galaxy is the main member in a group of
  eight; a luminosity excess (halo) is detected from \mbox{$ r
  \sim 5 $ Kpc} outwards (figure \ref{fig:parametros1}).

\item NGC~7177: the surface brightness profile lies below the
  $R^{1/4}$ law for $ r < 3.8''$; this behaviour cannot be explained
  in terms of a seeing effect: we have computed the expected light distribution
  for a pure  $R^{1/4}$ bulge with $r_e  \approx 6.5\arcsec $  convolved 
 with the same seeing PSF (about
  1.2\arcsec FWHM) finding that the expected profile flattening begins
  about $ r \approx 0.65''$. Therefore, the observed flattening seems to be 
  an intrinsic bulge characteristic.  

\end{itemize}

\section{Discussion}

\subsection{Bulge and disc parameters of active and normal galaxies}

\subsubsection{The extended sample}

In order to perform a detailed study of the behaviour of bulge and disc 
parameters across all levels of nuclear activity, we have increased
our sample by compiling from the
literature a large number of morphological parameters belonging to
 galaxies contained in the 
HFS catalog. Our main source of data is the set of bulge and disc
parameters from  
Baggett et al. \shortcite{Baggett} (hereafter BBA) that comprises 659
galaxies, from which we have selected bulge and disc parameters for
the 234 objects contained in the
HFS catalog.

The extended sample contains 261 galaxies hosting all levels of nuclear 
activity. According to the classification of HFS the sample
 contains 22 Seyfert~1,
16 Seyfert~2, 90 LINERs, 108 starburst and HII galaxies and 25 normal
galaxies. Therefore, there is an
excess of low-activity nuclei. 

BBA parameters were derived using profiles obtained from major axis cuts of 
two-dimensional digital images created from photographic plates (photographic
$V$ band), from the PANBG catalog \cite{Kodaira1990}. The field size is large 
but the spatial resolution is quite poor ($1''/pixel$), as is the mean seeing,
about 4\arcsec FWHM. On the other hand, our results were obtained from azimuthally 
averaged profiles of CCD images with good spatial resolution ($0.3''/pixel$), 
much better mean seeing ($\sim 1.45''$ FWHM) but reduced field of view, 
limiting the maximum profile radius to about 80\arcsec. Therefore, we can expect 
from BBA data a better
disc sampling while our profiles would provide more accurate bulge results.

\begin{figure*}
\begin{minipage}{120mm}
\centering
\psfig{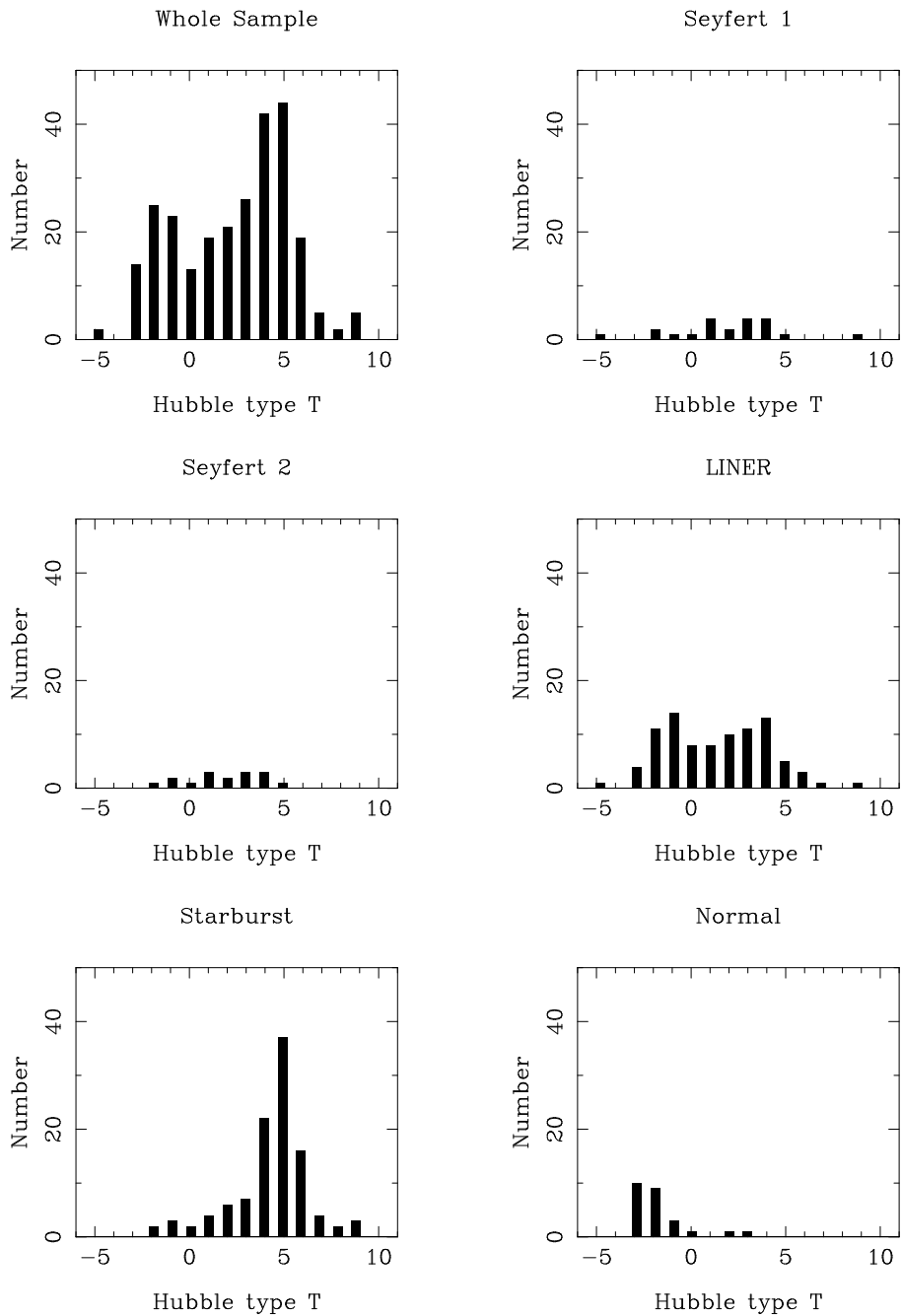}
\caption{Distribution of morphological Hubble types of the extended sample galaxies}
\label{morpho_all}
\end{minipage}
\end{figure*}

\begin{figure}
\hspace{1cm}
\psfig{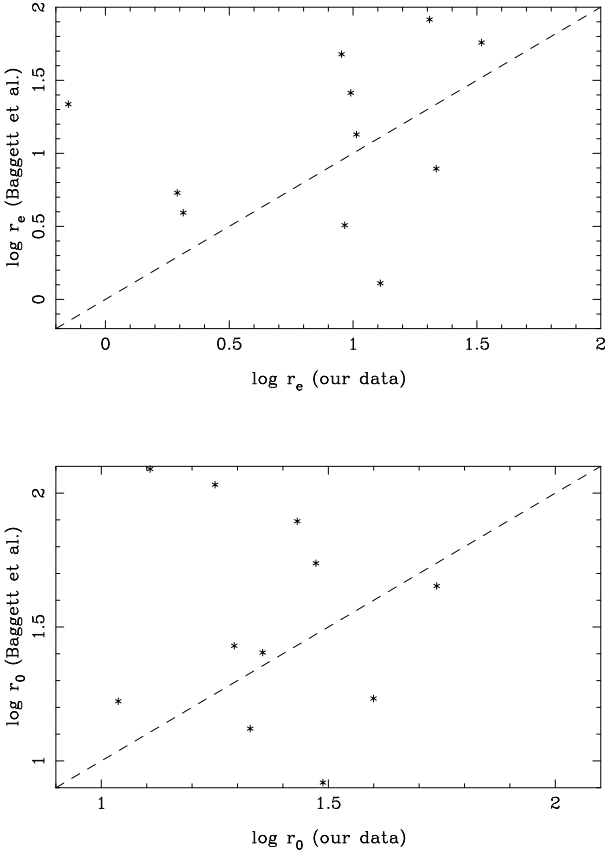}
\caption{Comparison between disc (r$_0$) and bulge parameters (r$_e$)
from our data and BBA data.}
\label{fig:comparison}
\end{figure}

Due to the differences shown above, in some cases large discrepancies  
between our results and those from BBA appear, as shown in figure
\ref{fig:comparison}.

The sample morphological distribution is shown in figure \ref{morpho_all}. 
The results of a $\chi^{2}$ test on the distribution for the different
nuclear types is presented in table \ref{chi_square}. Seyfert~1 and 
Seyfert~2 galaxies are drawn from the same
morphological distribution at 99\% confidence level; Seyfert~2 and
LINERs are also drawn from the same distribution at nearly 99\%
confidence level; Seyfert~1 and
LINERs follow the same distribution at a lower confidence level
($\simeq$ 79\%); on the other hand, the remaining nuclear types 
present clearly different
morphological type distributions. This should be taken into account
when analysing the distribution of bulge and disc parameters.

\begin{table}
\centering
\caption{Chi-square test on the morphological distribution for the
  different nuclear types}
\label{chi_square}
\begin{tabular}{lcc}
\hline
Comparison types & $\chi^{2}$ & Significance \\
\hline
S1 vs. S2              &  2.408 & 0.992 \\
S1 vs. LINER           &  8.643 & 0.799 \\
S1 vs. starburst       & 29.920 & 0.005 \\
S1 vs. normal          & 29.182 & 0.002 \\
S2 vs. LINER           &  4.314 & 0.987 \\
S2 vs. starburst       & 22.016 & 0.037 \\
S2 vs. normal          & 24.120 & 0.004 \\
LINER vs. starburst    & 64.457 & 0.000 \\
LINER vs. normal       & 38.634 & 0.000 \\
Starburst vs. normal   & 96.736 & 0.000 \\
\hline
\end{tabular}
\end{table}

According to HFS, all spiral galaxies show nuclear emission lines. Normal
galaxies (with no emission lines in the nuclear spectrum) are only observed in
the earliest Hubble types ($T\leq 0$). Most AGNs reside in early-type galaxies
(E-Sbc) while starbursts are generally observed in later type galaxies.

As an independent comparison sample, we have added the 
$V$ and $I$ data from Xanthopoulos 
\shortcite{MNRAS280:Xanthopoulos} and from Mediavilla et al.
\shortcite{ApSS157:Mediavilla}.  The former sample comprises
Seyfert type 1 and 2 galaxies from  V\' eron-Cetty \& 
V\' eron \shortcite{Veron}, with declination between -55$^{\circ}$ and 
+8$^{\circ}$ and redshift $z \leq 0.043$. Hubble types range from -2
(S0) to 5 (Sc); the latter also comprises Seyfert~1 and 2 galaxies
from Adams \shortcite{Adams77}, with redshift $z \leq 0.08$. This
sample includes one E galaxy, one N-type galaxy and several disc
galaxies with Hubble type ranging from -2 (S0) to 4 (Sbc).
This complementary data set comprises 21
Seyfert~1 and 12 Seyfert~2 galaxies.

\subsubsection{Disc and bulge parameters}
\label{disc_bulge_parameters}

We have performed a statistical analysis on the individual morphological
parameter distributions by means of a series of Kolmogorov-Smirnov
tests. The results are shown in table \ref{ks}. 
It should be recalled that the distribution of morphological
types accross Seyfert~1, Seyfert~2 and LINER galaxies is quite
homogeneous, but that is not true when compared with starbursts and
normal galaxies (see table \ref{chi_square}). Seyfert~1 and Seyfert~2 
effective radii and scale lengths are
drawn from the same distribution ({\em null hypothesis}) at about 97\%
and 91\% confidence level, respectively. On the other hand, while
bulges of Seyfert~1 galaxies and LINERs follow the same distribution
at about 96\% confidence level, disc distributions are most likely
different (null hypothesis is rejected at 48\% confidence level). 
 Seyfert~2 and LINER
bulge effective radii are drawn from the same distribution only 
at less than 70\% level, while disc scale lengths follow the same
distribution at about 90\% level. When comparing active
(i.e. Seyfert~1, Seyfert~2 and LINER) galaxies with non-active
(i.e. starburst and normal) galaxies, distribution of morphological
parameters are generally different (with some exceptions, e.g. disc scale
lengths of Seyfert~2 and normal galaxies). As a conclusion, we cannot
claim that the individual morphological parameters of active and 
non-active galaxies follow the same statistical distribution, 
as other authors do (e.g. Virani et al. 2000).

\begin{table*}
\caption{Kolmogorov-Smirnov test on the distribution of morphological 
parameters}
\label{ks}
\begin{tabular}{lcccccccccc}
\hline
Comparison & \multicolumn{2}{c}{$\mu_e$} & \multicolumn{2}{c}{$r_e$} &
\multicolumn{2}{c}{$\mu_0$} &  \multicolumn{2}{c}{$r_0$} &
\multicolumn{2}{c}{$B/D$}\\
types & K-S & Significance &  K-S & Significance &
K-S & Significance &  K-S & Significance &   K-S & Significance \\
\hline
S1 vs. S2              & 0.227 & 0.725 & 0.148 & 0.988 & 0.317 & 0.440 & 0.167 & 0.985 & 0.245 & 0.786 \\
S1 vs. LINER           & 0.127 & 0.941 & 0.121 & 0.961 & 0.188 & 0.627 & 0.188 & 0.627 & 0.233 & 0.356 \\
S1 vs. starburst       & 0.202 & 0.489 & 0.256 & 0.216 & 0.211 & 0.444 & 0.345 & 0.037 & 0.300 & 0.122 \\
S1 vs. normal          & 0.280 & 0.328 & 0.295 & 0.269 & 0.200 & 0.819 & 0.250 & 0.560 & 0.647 & 0.001 \\
S2 vs. LINER           & 0.209 & 0.596 & 0.196 & 0.677 & 0.267 & 0.448 & 0.171 & 0.921 & 0.297 & 0.366 \\
S2 vs. starburst       & 0.253 & 0.366 & 0.275 & 0.271 & 0.311 & 0.248 & 0.300 & 0.287 & 0.445 & 0.046 \\
S2 vs. normal          & 0.292 & 0.388 & 0.312 & 0.306 & 0.433 & 0.120 & 0.200 & 0.925 & 0.455 & 0.112 \\
LINER vs. starburst    & 0.219 & 0.043 & 0.266 & 0.007 & 0.121 & 0.522 & 0.185 & 0.092 & 0.402 & 0.000 \\
LINER vs. normal       & 0.303 & 0.063 & 0.283 & 0.099 & 0.288 & 0.142 & 0.162 & 0.792 & 0.467 & 0.003 \\
Starburst vs. normal   & 0.427 & 0.003 & 0.250 & 0.206 & 0.239 & 0.293 & 0.181 & 0.646 & 0.847 & 0.000 \\

\hline
\end{tabular}
\end{table*}

\begin{figure*}

\hspace{1cm}
\psfig{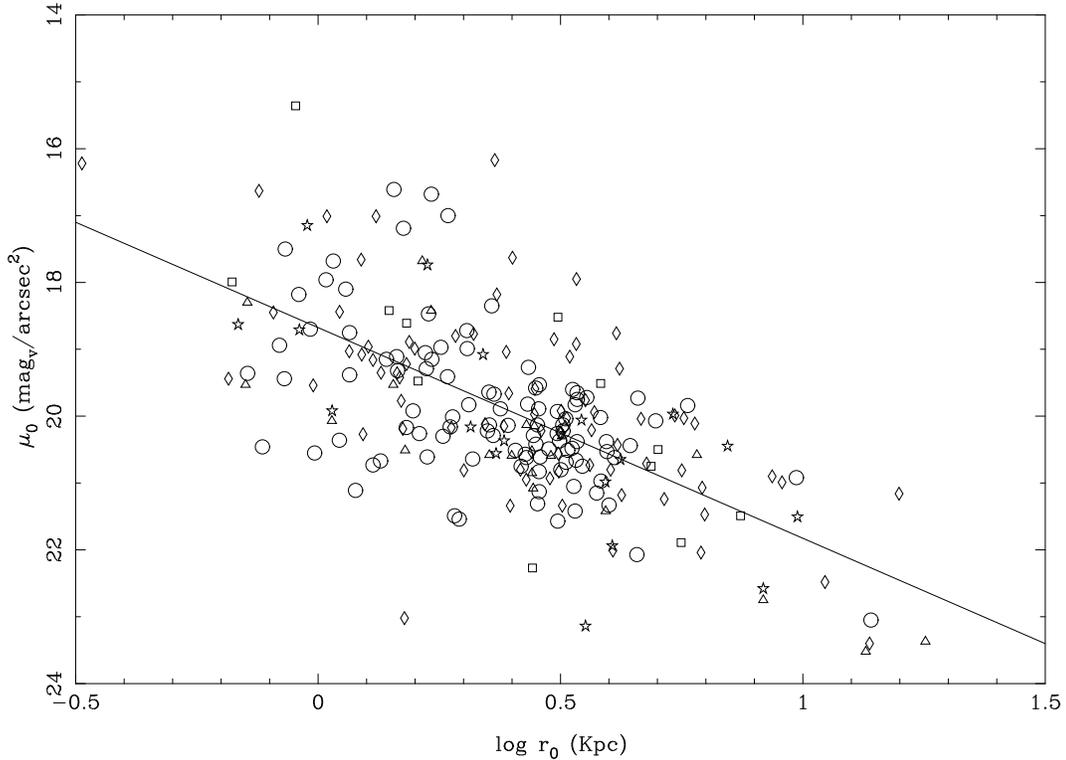}
\caption{Disc parameters  $\mu_0$ and $\log r_0$
  (mag arcsec$^{-2}$, radii in Kpc). Seyfert~1 galaxies are represented  as
star symbols ($\star$) , Seyfert~2 as squares ($\Box$), LINERs as
diamonds ($\diamondsuit$), Starbursts as circles ($\circ$) and normal
galaxies as triangles ($\triangle$). The solid line shown is the
result of a linear regression. A weak linear correlation is observed (the
correlation coefficient is only about 0.65) }
\label{fig:disk_parameters}
\end{figure*}

Perhaps more interesting than the statistics shown above, 
the distribution of morphological parameters in the ($\mu_0$, $\log
r_0$) and ( $\mu_e$, $\log r_e$) planes can be studied. 
Disc parameters are strongly
clustered (perhaps a selection
effect due to the luminosity-limited sample)
but there is only a weak linear correlation in the ($\mu_0$, $\log
r_0$) plane, as shown in figure \ref{fig:disk_parameters}.

On the other hand, bulge parameters show a strong linear correlation
that constitutes the projection in the ($\mu_e$, $\log
r_e$) parameter subspace  of the ``fundamental plane'' of ellipticals
and bulges of spiral and lenticular galaxies. 

\begin{figure*}
\hspace{1cm}
\psfig{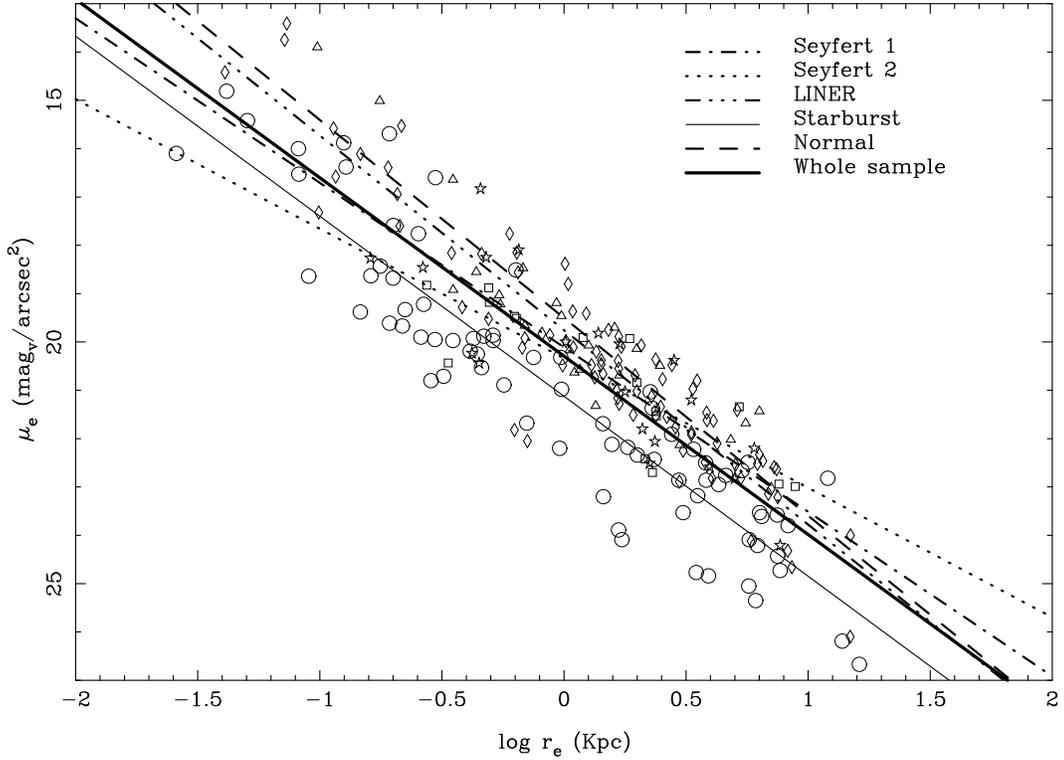}
\caption{Bulge parameters  $\mu_e$ and $\log r_e$
  (mag arcsec$^{-2}$, radii in Kpc). Symbols as in figure 
\ref{fig:disk_parameters}. Regression lines have been plotted for all
nuclear types.}
\label{fig:bulge_parameters}
\end{figure*}

\begin{figure*}
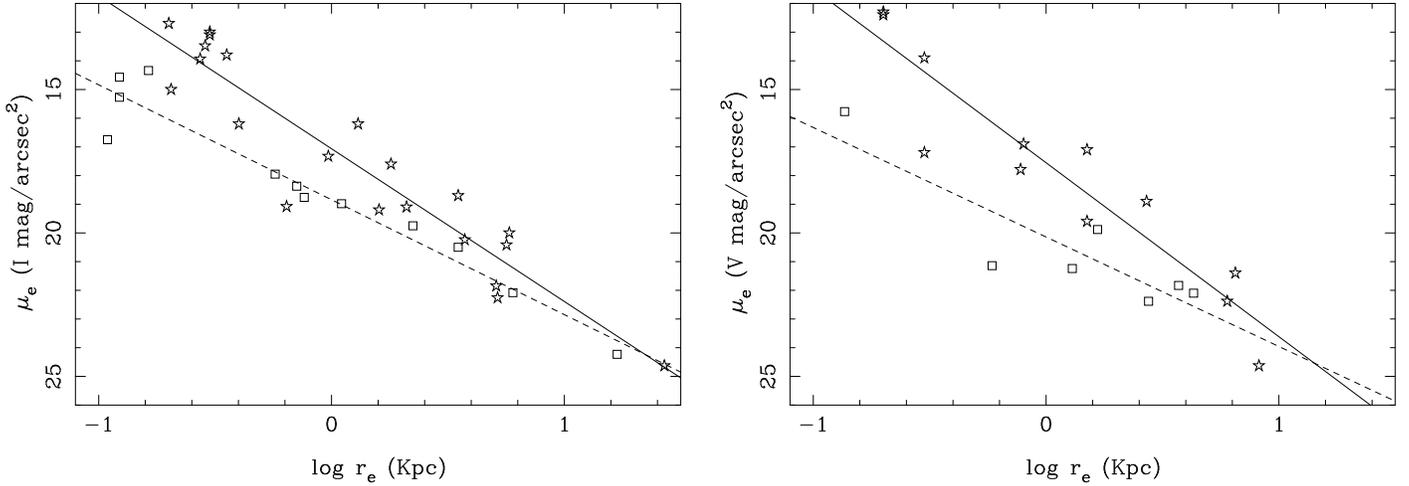

\hbox{
\hspace{-3mm}
\psfig{figure=comparison_sample_i.ps,width=90mm,angle=270}
\hspace{3mm}
\psfig{figure=comparison_sample_v.ps,width=90mm,angle=270}
}
\caption{Bulge parameters  $\mu_e$ and $\log r_e$
  (mag arcsec$^{-2}$, radii in Kpc) for the comparison sample in the
  $I$ (left) and $V$ bands. Symbols as in figure 
\ref{fig:disk_parameters}. The solid line corresponds to the linear
  regression for Seyfert~1 galaxies and the dashed line to Seyfert~2
  galaxies. Differences in the slope of both distributions are
  $\simeq 2.4 \sigma $ in the $I$ band and $\simeq 1.9 \sigma $ in the
  $V$ band.}
\label{comparison_sample}
\end{figure*}

We have performed a fit of the relation between bulge parameters $\mu_e$ vs. $\log r_e$
by means of linear regression for the different nuclear types. The
results are shown in figure \ref{fig:bulge_parameters}.  We can observe
that: 

\begin{enumerate}

\item The spheroidal component of starbursts constitutes a
lower luminosity class. It is clear in figure
\ref{fig:bulge_parameters} that starbursts follow a distribution that
lies parallel and below the overall sample. This could be related to
the fact that starburst galaxies tend to be late-type
galaxies, as is seen in figure \ref{morpho_all}.

\item  LINERs and normal galaxies follow very similar
distributions (differences in slope are smaller than $0.2\sigma$ and
$\sim \sigma$ in zero point).

\item We can observe differences in the slope of the distributions of
  Seyfert~1, Seyfert~2, LINERs and normal galaxies. When comparing
  Seyfert~1 with LINERs and normal galaxies, the differences are of
the order 
  $\sigma$ and can be due to statistical errors. Differences
  between the Seyfert~2 distribution and that of LINERs and 
  normal galaxies are more significative, $\gsim 2.5 \sigma$ level. On
  the other hand, there is a difference at  $\sim \sigma$ level
in the slope of the distribution of  Seyfert~1 and Seyfert~2 that 
can be also attributed to
statistical uncertainties. Nevertheless, equivalent fits performed in both $V$
and $I$ bands in our comparison sample lead to a very similar result
at $\sim 2 \sigma$ significance level (see figure 
\ref{comparison_sample}). 
Therefore, we have observed that the different classes of galaxies
(Seyfert~1, Seyfert~2, LINERs and normal galaxies) show different
bulge distributions in the ($\mu_e$, $\log r_e$) plane. While
sometimes this difference has little statistical significance, in the
case of Seyfert~2 galaxies it is more important, since it cannot be
solely attributed to statistical effects. If real, the observed
distribution difference could imply families of bulges with {\em
  different} physical properties, as proposed for bright and faint
cluster ellipticals \cite{Hoessel}.  Capaccioli et
al. \shortcite{Capaccioli92} also describe two physically different
families of
 ellipticals, early type dwarfs and bulges of SOs and spirals, each family 
described by a different
relation in the ($\mu_e$, $\log r_e$) plane. One of the families
comprises the cluster brightest  cD and quasar-hosting
galaxies, while the other comprises normal ellipticals, fainter
bulges and early type dwarfs. According to these authors, the latter objects are in the
latest stages of dissipative collapse, while the former ones have
experienced merging and accretion processes. 

The stellar kinematics in the nuclei of Seyfert galaxies both of types 1
and and 2 seems to be similar to that in normal galaxies, with active
and non-active galaxies following the same Faber-Jackson relation
(Terlevich, Diaz \& Terlevich 1990; Nelson \& Whittle 1996). Further
evidences on the similarity of stellar kinematics in active and normal
galaxies has been provided by Ferrarese
et~al. \shortcite{Ferrarese01}. They studied the $M_{\bullet} - \sigma$
relation for AGNs, an empirical correlation between the central black hole
mass and the stellar velocity dispersion, finding a good agreement
between black hole masses derived from reverberation mapping and from
the $M_{\bullet} - \sigma$ relation as derived from normal and weakly
active galaxies, indicating a common relationship between active and
quiescent black holes and their host galaxy environment. Regarding gas
motions the relation between the velocity dispersion of the gas in the
NLR and the stellar velocity dispersion and for galaxies of Seyfert
types 1 and 2 shows a large scatter although clustered around the 1:1
line (Jim\' enez-Benito et~al. 2000). This suggests that gas motions are mainly
controlled by the mass of the galactic bulge along with other agents, like
shock waves that may be produced by the interaction of jets with ambient
gas, or tidal interactions with companions, providing an aditional line
broadening. Yet, the kinematics of the two main types of Seyfert galaxies
show some differences that need to be explained: a ratio between gas and
stellar velocity dispersions larger than one is common in Seyfert type 2
nuclei, while some Seyfert type 1 nuclei show emission lines which are
narrower than the stellar absorption lines.

\end{enumerate}

\subsubsection{Type II discs}
\label{TypeII_discs}

As shown in section \ref{notes} we have found type II discs in three
Seyfert galaxies and in one LINER from our original sample. 
Since no type II disc was detected
in non-active (i.e. starburst or normal) galaxies, we have investigated
the possible connection between nuclear activity and the 
existence of type II discs.
The BBA sample parameters were derived using the
modified Kormendy function (equation \ref{eq:diskcut}) and a large
fraction of the objects ( $\simeq$ 50\%) 
presents a type II disc.

Tables \ref{type2_nt} and \ref{type2_mt} show the statistics of type
II discs in the extended sample. As shown in table \ref{type2_nt}, 
we don't find correlation between the existence of type II discs and nuclear 
activity. In fact, the highest fraction of type II discs is found in
normal galaxies. Nevertheless, we do find a greater incidence 
of type II discs in earlier 
morphological types (\mbox{T = -3} to \mbox{T = 0}) as shown in 
table \ref{type2_mt}. 

\begin{table}
\caption{Type II discs vs. nuclear type}
\label{type2_nt}
\begin{tabular}{lccc}
\hline
Nuclear type & No. of objects & No. of objects & Ratio (\%)\\
 & with disc fit & with type II disc  & \\
 \hline
Seyfert 1 & 19 & 6 & 31.6 \\
Seyfert 2 & 12 & 4 & 33.3 \\
LINER & 80 & 31 & 38.7 \\
Starburst & 103 & 24 & 23.3\\
Normal & 20 & 8 & 40.0 \\
\hline
\end{tabular}
\end{table}

\begin{table}
\caption{Type II discs vs. Hubble type}
\label{type2_mt}
\begin{tabular}{cccc}
\hline
Hubble type   & No. of objects & No. of objets &  Ratio (\%)\\
 T &  with disc fit  & with type II disc & \\
 \hline
-3 & 11 & 6 & 54.5 \\
-2 & 21 & 9 & 42.9 \\
-1 & 21 & 11 & 52.4 \\
0 & 12 & 6 & 50.0 \\
1 & 16 & 3 & 18.7 \\
2 & 19 & 3 & 15.8 \\
3 & 25 & 7 & 28.0 \\
4 & 38 & 10 & 26.3 \\
5 & 41 & 14 & 34.1 \\
6 & 19 & 4 & 21.0 \\
\hline
\end{tabular}
\end{table}

\subsubsection{$B/D$ relation}
\label{b_d_relation}

We have derived the $B/D$ relation by means of equation
\ref{bd_2}. As comparison values, we have used the mean values
obtained by Simien \& de Vaucouleurs \shortcite{Simien} (hereafter
SDV); these values served as a reference data set,
rejecting all morphological parameters leading to $B/D$ values
highly deviated from SDV data. Nevertheless, it is worth mentioning
that SDV $B/D$ values present a large scatter, mainly attributed to
errors in the surface photometry and in the decomposition process and,
to a lesser extent, to morphological classification errors.

\begin{figure*}
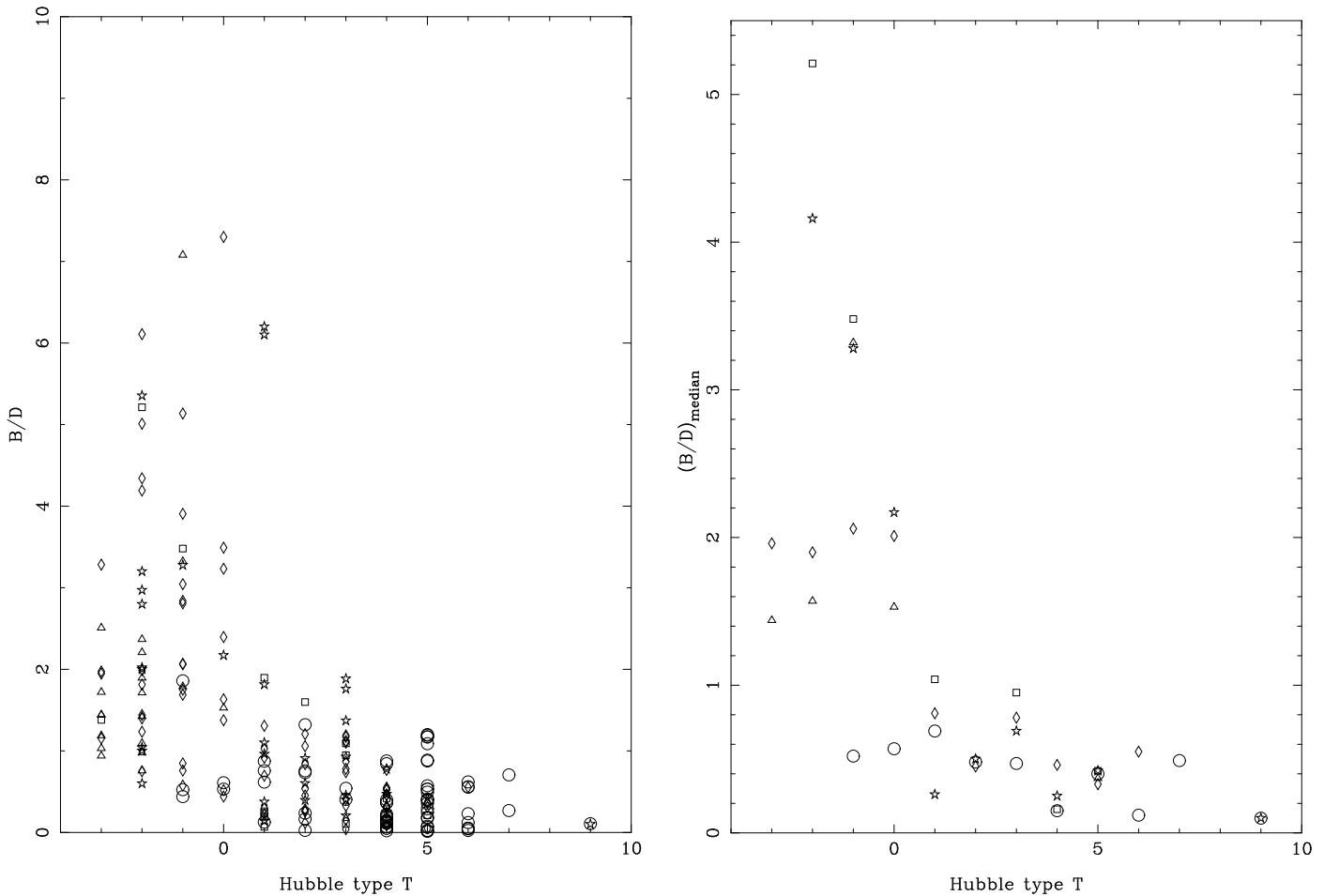

\hspace{1cm}
\hbox{
\hspace{-3mm}
\psfig{figure=bd.ps,width=90mm}
\hspace{3mm}
\psfig{figure=bd_mean.ps,width=90mm}
}
\caption{Distribution of $B/D$ relation (left pannel) and median value
  $(B/D)_{median}$ (right pannel) for the different
  morphological types. Symbols as in figure 
\ref{fig:disk_parameters}.}
\label{fig:bd}
\end{figure*}

The distribution of  $B/D$ values is shown in figure \ref{fig:bd}
(left pannel). As
can be observed in table \ref{ks}, the $B/D$ relation values for
 Seyfert~1 and Sefyert~2 galaxies are drawn from the same distribution 
and 86\% confidence level. On the other hand, Seyfert and LINER
 $B/D$ distributions are most likely different since null hypothesis
is rejected at about 26\% and 37\% confidence level for Seyfert~1 and
 Seyfert~2, respectively.

As can be observed in figure \ref{fig:bd} and in the mean deviation estimators
shown in table \ref{bd_mt},  $B/D$ values spread over a large
range,  specially for the earliest morphological types (T $<$
 0). In order to avoid the effects of outliers,
 we have used the median $(B/D)_{median}$ rather than the mean 
as a robust average estimator.
As shown in table \ref{bd_mt} and in the right pannel of figure
\ref{fig:bd}, a descending trend is observed in $(B/D)_{median}$ with
the Hubble type T, as expected from the definition of the latter as a
sequence of decreasing importance of the spheroidal component with
respect to the disc. This trend is also reflected in SDV data.
Our median values are generally higher than
those derived by SDV (see table \ref{bd_mt}); this can be at least
partially explained by the fact that  
we allow for the existence of type II discs (not considered by SDV);
on the other hand, SDV values are derived from the $B$ band while ours
are computed from $V$ band surface profiles. From 
table \ref{bd_mt} and figure \ref{fig:bd} we can observe that:

\begin{enumerate}
\item The distribution of $(B/D)_{median}$ seems to be generally independent of the
  nuclear type.

\item Nevertheless, for the earliest morphological types (T $<$ 0), Seyfert 
  galaxies tend to have a higher $(B/D)_{median}$ relation (greater relative
  importance of the spheroidal component) than the remaining types.

\item  Starburst galaxies show smaller $(B/D)_{median}$ values than the
  remaining nuclear species for the
  earliest Hubble types (T $<$ 0) and approximately constant
between \mbox{T = -1} and  \mbox{ T = 3}. 

\item From \mbox{T = 1} onwards, $(B/D)_{median}$ is quite uniform across all nuclear
  types.

\end{enumerate}

\begin{table}
\caption{Median $B/D$ relation vs. morphological type. SDV mean $B/D$ values are
  shown for comparison}
\label{bd_mt}
\begin{tabular}{cccccc}
\hline
Hubble  & No. of & $(B/D)_{median}$ & mean & $\langle B/D \rangle$
& mean \\
type  T & objets & & dev. & (SDV) & dev. (SDV) \\
\hline
-3 & 11 & 1.44 & 0.52 & 1.43 & 0.42 \\
-2 & 21 & 1.89 & 1.29 & 1.33 & 0.24 \\
-1 & 20 & 2.06 & 1.23 &1.27 & 0.32 \\
0 & 12 & 1.58 & 1.27 & 1.02 & 0.13 \\
1 & 16 & 0.66 & 0.45 & 0.55 & 0.32 \\
2 & 18 & 0.49 & 0.36 & 0.48 & 0.10 \\
3 & 20 & 0.76 & 0.38 & 0.29 & 0.05 \\
4 & 33 & 0.25 & 0.19 & 0.22 & 0.06 \\
5 & 29 & 0.40 & 0.29 & 0.11 & 0.02 \\
6 & 10 & 0.18 & 0.22 & 0.04 & 0.01\\
7 & 2 & 0.49 & 0.22 & 0.02  & 0.01 \\
8 & - & -  & -  & - & - \\
9 & 2 & 0.10 & 0.003 & - & - \\
\hline
\end{tabular}
\end{table}

\subsection{Radial colour distribution}
\label{radial_colour_distribution}

Yee \shortcite{ApJ272:Yee} performed Gunn {\em
  r, g, v } surface photometry on a sample of twenty Seyfert galaxies
  from the Markarian catalog concluding that the colours of the
  underlying galaxies are comparable to those of normal spirals in the
  Sa-Sbc range, but the nuclear colours are bluer in 
Seyfert~1 galaxies than in Seyfert~2 objects. Xanthopoulos 
\shortcite{MNRAS280:Xanthopoulos} performed aperture photometry on a
  sample of 27 Seyfert~1 and Seyfert~2 galaxies using 
2.5, 5, 10 and 20~Kpc apertures.
No differences (within
  the measurement uncertainties) were found in the 
\mbox{$V-R$}, \mbox{$R-I$} and \mbox{$V-I$} colours in the 10 and
  20~Kpc apertures characterizing the disc. 
In the 2.5~Kpc and 5~Kpc apertures, however,  the
  \mbox{$R-I$}, and, to a lower degree, the \mbox{$V-I$} colours
  appear bluer in Seyfert~1 galaxies than in Seyfert~2.

It should be taken into account that AGN aperture photometry is contaminated
by nuclear emission lines and it is expected that this effect would be more
noticeable in the smaller diaphragms. This is specially true for the 
\mbox{$R-I$} colour in Seyfert~1 galaxies, since the $R$ band contains the
broad component of \Ha .  \mbox{$R-I$} colours are then expected to be bluer
in Seyfert~1 galaxies than in Seyfert~2 ones. Since we are interested in 
characterising the underlying galaxy colours, we would need to 
parametrise them avoiding the nuclear emission lines. 
To this end, we have used the radial
colour distributions presented in Paper I, 
analysing the  \mbox{$R-I$} and \mbox{$V-I$} colour profiles. While radial
distributions of bulges and discs generally become bluer outward (Balcells \&
Peletier, 1994), gradients are small enough that it is possible to
assign representative values for the colour of each component. For
instance, Peletier \& Balcells \shortcite{PB96} selected as 
characteristic of bulges the
colour at 0.5 $\times$ r$_e$ or at 5 arcsec, whichever is larger, and for
discs the colours at 2 $\times$ r$_0$. In order to minimise the risk of
nuclear contamination at small radii in bulges and 
too high colour errors at large radial distances in discs, we
have decided to select as representative the colours at effective 
radius (r$_e$)
for bulges and those at scale length (r$_0$) for discs.

\begin{figure*}
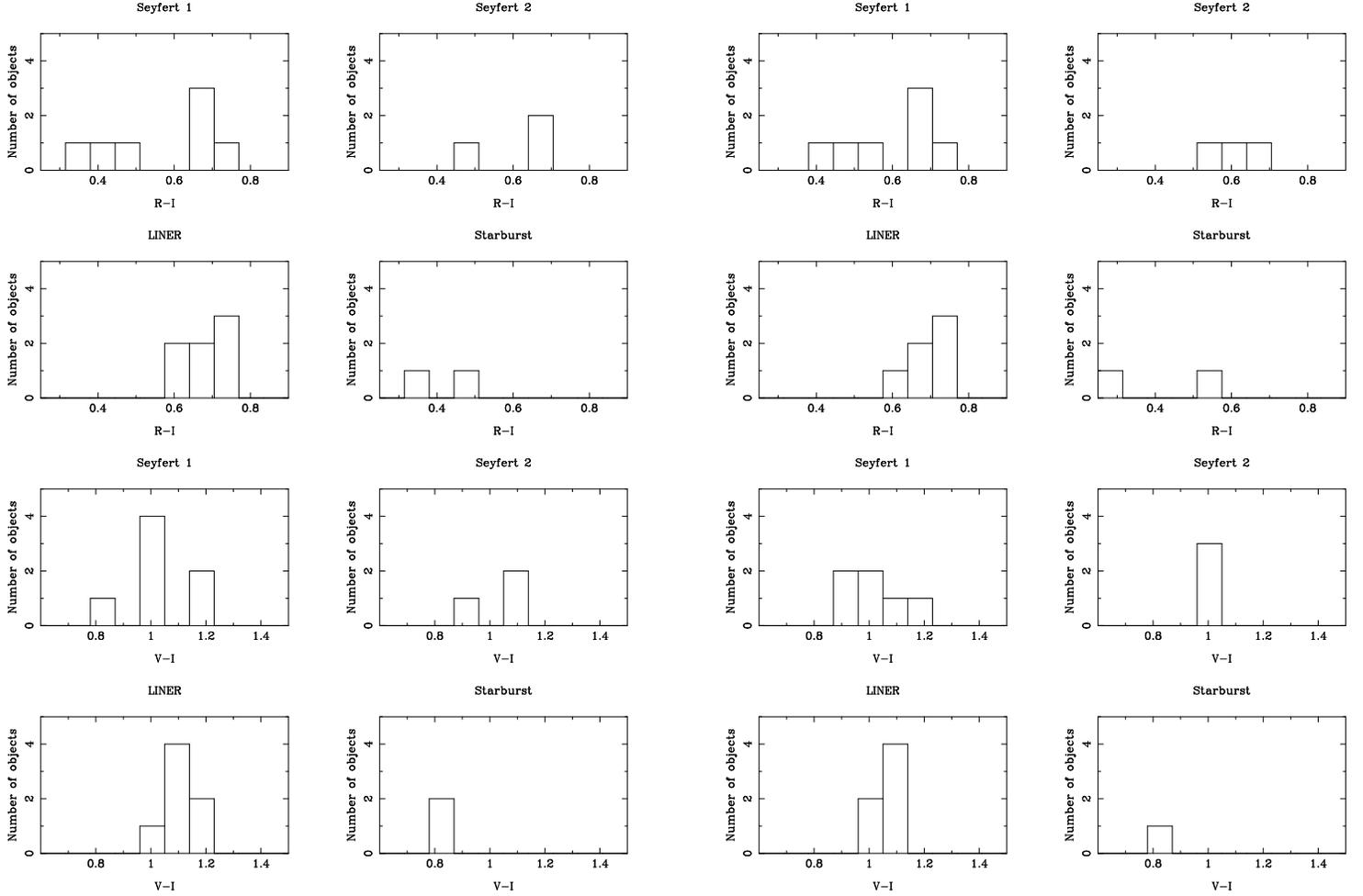

\hspace{1cm}
\hbox{
\hspace{-12mm}
\psfig{figure=hist_color_bulge_r_eff.ps,width=90mm}
\hspace{12mm}
\psfig{figure=hist_color_disc_r_0.ps,width=90mm}
}
\caption{Distribution of bulge (left) and disc characteristic colours}
\label{hist_colours}
\end{figure*}

The results
are somewhat qualitative, since statistics are not reliable for such a
small sample. As shown in figure \ref{hist_colours}, we have found 
that the range of characteristic bulge
and disc colours is comparable in all nuclear types, albeit mean
bulge and disc characteristic colours are bluer in Seyfert
galaxies than in LINERs. Seyfert~1 bulge and disc characteristic
colours are somewhat bluer than those of Seyfert~2. Nevertheless, it
should be taken into account that the bluest Seyfert~1 observed bulge
colours are almost certainly contaminated by either the NLR component (NGC~3227) or by
circumnuclear star-forming rings (NGC~3982 and NGC~7469)

\begin{figure*}
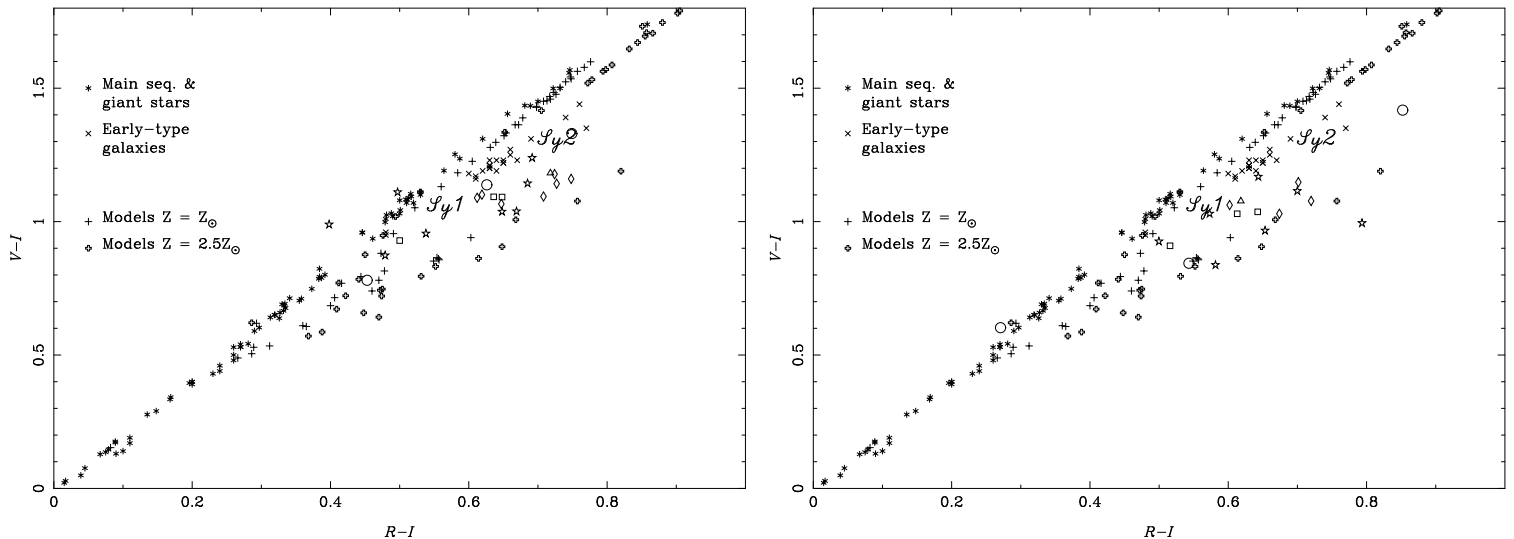

\hspace{1cm}
\hbox{
\hspace{-12mm}
\psfig{figure=color_bulge_compar1.ps,angle=270,width=98mm}
\hspace{1mm}
\psfig{figure=color_disk_compar1.ps,angle=270,width=98mm}
}
\caption{Bulge (left) and disc (right) characteristic colours compared with  
photoelectric photometry of stars and early-type galaxies and with
population synthesis-derived colours. Symbols as in figure 
\ref{fig:disk_parameters}. The points labelled as {\it Sy1}
and {\it Sy2} correspond to the mean 
Seyfert~1 and Seyfert~2 colours in the 5 Kpc aperture from
Xanthopoulos (1996).}
\label{fig:color_1}
\end{figure*}

As a first comparison, we have represented in figure \ref{fig:color_1} 
our bulge and disc colours
along with photoelectric photometry values of standard stars from 
Moreno \& Carrasco \shortcite{AA65:Moreno} and Graham
\shortcite{PASP94:Graham} and aperture photometry of early-type galaxies
from Poulain \shortcite{Poulain}. We have also included colours from
population synthesis models computed by Bressan et al. \shortcite{Bressan} at 
 $Z = Z_{\odot}$ and $Z = 2.5 Z_{\odot}$ and a large range of ages
(from $6.3 \times 10^6$ to $19.95 \times
  10^9$ years). There is an acceptable agreement between our colours
and those of the comparison sample: they are comprised between those
derived from aperture photometry of early-type galaxies (whose colours
are generally redder) and the population synthesis models.

\begin{figure*}
\psfig{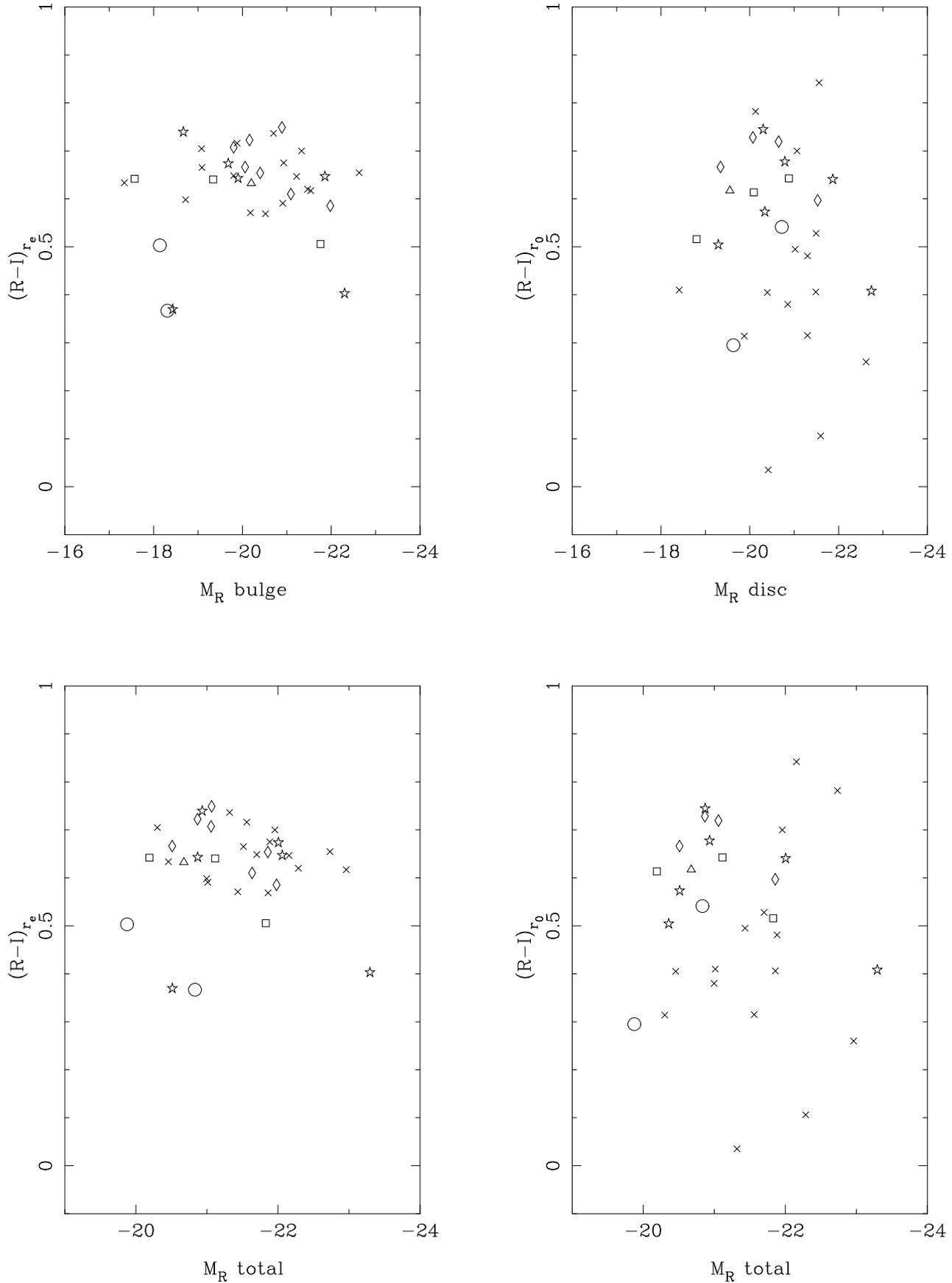}
\caption{Sample bulge (left diagrams) and disc (right diagrams) characteristic colours 
(symbols as in figure 
\ref{fig:disk_parameters}) plotted against bulge, disc and total
absolute magnitudes. The comparison sample ($\times$ symbols)
comprises 18 early-type galaxies from Balcells \& Peletier
\shortcite{BP94} and Peletier \& Balcells \shortcite{PB96}.}
\label{fig:ri_mabs}
\end{figure*}

A further step is the comparison of our characteristic colours with those of normal
galaxies with similar morphological type. 
To this end, we have selected a reference sample of 18 early-type
galaxies (Balcells \& Peletier, 1994; Peletier \& Balcells,
1996. Those studies will be referenced hereafter as BP).
Three of them
are catalogued as normal galaxies in HFS, while four of the galaxies
have been labelled as transition types or LINERs. The remaining eleven
objets are not included in the HFS sample, but have been checked in
NED and in V\' eron-Cetty \& V\' eron \shortcite{Veron} 
for the absence of active nucleus.

Characteristic colours from BP have been derived from colour gradients
rather than from the colour profiles themselves. These authors find that
bulge colours are predominantly bluer than those of elliptical
galaxies, but both bulges and ellipticals follow the same sequence in
the colour-colour diagram. They conclude that stellar populations of
bulges must be similar to those of elliptical galaxies, but the
former, as a class, have lower metallicities and larger spread in ages
than giant ellipticals. Visvanathan \& Sandage \shortcite{Visvanathan}
found that ellipticals and S0 galaxies follow a colour-magnitude
relation in which brighter galaxies are redder; BP also tried  
to correlate bulge colours and
global galaxy parameters. They find that the (weak) correlation
between colours and absolute magnitude improves slightly when the
total galaxy luminosity is used instead of the bulge luminosity
alone. This might indicate that the total galaxy potential, and not
the bulge alone, determines the chemical enrichment of the
bulge. Moreover, they find that bulge and disc colours derived from
profiles are very similar, suggesting that the bulk of the inner disc
and bulge stars are essentially coeval. At most, disc stars are 2-3
Gyr younger than bulge stars.

Figure \ref{fig:ri_mabs} shows the relationship between our
characteristic \mbox{$R-I$} colours and the bulge, disc and total
galaxy luminosity. BP data have also been plotted in order to compare
the control sample colours with ours. We find that:
\begin{enumerate}
\item When comparing our sample with the control sample, we find a
very similar colour range; this is true for both bulge and disc
components and may suggest that 
the bulge and disc stellar populations (ages and metallicities)
are comparable in normal and active galaxies.

\item We confirm the result from BP that bulges and disc
characteristic colours derived from profiles are very similar, though
there is a wider range of disc characteristic colours.

\item We are not able to confirm a correlation between bulge colours
and bulge luminosity. Moreover, this situation does not improve when
using the total galaxy luminosity instead of the bulge luminosity
alone.

\end{enumerate}

\section{Conclusions}
\label{conclusions}

We have investigated the main structural
properties derived from {\em VRI} and \Ha\ surface photometry of galaxies hosting
nuclear emission-line regions (including Seyfert~1, Seyfert~2, LINER and
starburst galaxies) as compared with normal galaxies. Our original sample
is thoroughly described in Paper I and comprises 21 active galaxies, 3
starbursts and 1 normal galaxy. For our investigation on bulge and disc
parameter distributions, we have extended the sample up to 261 objects hosting all
levels of nuclear activity (ranging from normal to Seyfert~1 galaxies). Our
study has found several differences between the structural parameters of
active and non-active galaxies. From
the statistical analysis performed in section \ref{disc_bulge_parameters}, we
cannot claim that the individual morphological parameters of active and
non-active galaxies follow the same statistical distribution. On the other
hand, we have studied the distribution of morphological parameters in the 
($\mu_0$, $\log r_0$) and ( $\mu_e$, $\log r_e$) planes. 
Disc parameters are found strongly clustered (perhaps a selection effect) but
the linear correlation is weak. On the other hand, 
there is a strong linear correlation of bulge parameters. We have studied the
distributions accross the different nuclear types, finding that starbursts
constitute a lower luminosity bulge class. LINERs and normal galaxies follow
very similar distributions. Seyfert~1, Seyfert~2 and LINERs follow different
distributions at statistically significant level. These differences could
perhaps imply families of bulges with different physical properties. Type II
discs are analysed in section \ref{TypeII_discs}. We don't find a correlation
between type II discs and nuclear type but with Hubble type. 
$B/D$ relation is
studied in section \ref{b_d_relation}. We find that the distribution of
$(B/D)_{median}$ seems to be generally independent of the nuclear type, except
for the earliest ($T < 0$) Hubble types. Finally, bulge and disc
characteristic colours derived from radial profiles have been investigated in
section \ref{radial_colour_distribution}. We find that the range of
characteristic bulge and disc colours is comparable in all nuclear types,
though mean bulge and disc colours are bluer in Seyfert galaxies than in
LINERs. Nevertheless, this result is uncertain since bulge colours can be affected
by line emission contamination arising from the NLR and/or circumnuclear
star-forming regions. When comparing our characteristic bulge and disc colours with those of
the control sample of early-type galaxies,we find a
very similar colour range that may suggest that 
the bulge and disc stellar populations are comparable in normal and active galaxies.
On the other hand, we find that bulges and disc
characteristic colours derived from profiles are very similar.
We don't detect a correlation between bulge colours
and bulge or total galaxy luminosities.

\section*{Acknowledgments}

The JKT is operated on the island of La Palma by the Isaac Newton Group 
in the Spanish Observatorio del Roque de los Muchachos of the Instituto de
Astrof\'{\i}sica de Canarias. We would like to thank CAT for awarding
observing time. We also thank an anonymous referee for suggestions
that greatly improved the clarity of the paper.
This research has made use of the NASA/IPAC Extragalactic
Database (NED) which is operated by the Jet Propulsion Laboratory, California
Institute of Technology, under contract with the National Aeronautics 
and Space Administration.
M. S\'anchez would like to thank Isabel Casanova for her 
assistance entering and revising the large amount of data from  the literature.
This work has been partially supported by DGICYT project AYA-2000-0973.

\bsp

\label{lastpage}

\end{document}